\documentclass[fleqn,usenatbib]{mnras}

\usepackage[T1]{fontenc}

\usepackage{graphicx}	
\usepackage{amsmath}	
\usepackage{amssymb}	
\usepackage{multicol}

\usepackage{cancel}
\usepackage{newtxtext,newtxmath}

\title[Unveiling the nature of 11 DSFGs at $z\sim2$]{Unveiling the nature of 11 Dusty Star-Forming Galaxies at the peak of Cosmic Star Formation History}

\author[L. Pantoni et al.]{L. Pantoni,$^{1,2}$\thanks{E-mail: lpantoni@sissa.it}
A. Lapi,$^{1,2,3,4}$
M. Massardi,$^{5}$
D. Donevski,$^{1,3}$
A. Bressan,$^{1}$
L. Silva,$^{4}$
\newauthor{F. Pozzi,$^{6,7}$
C. Vignali,$^{6,7}$
M. Talia,$^{6,7}$
A. Cimatti,$^{6,8}$
T. Ronconi,$^{1,2,3}$
and L. Danese$^{1,3}$}
\\
$^{1}$SISSA - ISAS, Via Bonomea 265,  Trieste 34136, Italy\\
$^{2}$INFN - Sezione di Trieste, via Valerio 2, Trieste 34127, Italy\\
$^{3}$IFPU - Institute for fundamental physics of the Universe, Via Beirut 2, 34014 Trieste, Italy\\
$^{4}$INAF - Osservatorio Astronomico di Trieste, Via Giambattista Tiepolo, 11, 34131 Trieste, Italy\\
$^{5}$INAF - Istituto di Radioastronomia - Italian ARC, Via Piero Gobetti 101, I-40129 Bologna, Italy\\
$^{6}$DIFA - Dipartimento di Fisica e Astronomia, Universit\`{a} degli Studi di Bologna, Via Berti Pichat 6/2, I-40127 Bologna, Italy\\
$^{7}$INAF - Osservatorio di Astrofisica e Scienza dello Spazio di Bologna, Via Gobetti 93/3, I-40129 Bologna, Italy\\
$^{8}$INAF - Osservatorio Astrofisico di Arcetri, Largo E. Fermi, 50125, Firenze, Italy
}

\date{Accepted XXX. Received YYY; in original form ZZZ}

\pubyear{2021}

\begin{document}
\label{firstpage}
\pagerange{\pageref{firstpage}--\pageref{lastpage}}
\maketitle

\begin{abstract}
We present a panchromatic study of 11 (sub-)millimetre selected DSFGs with spectroscopically confirmed redshift ($1.5< z_{\rm spec}<3$) in the GOODS-S field, with the aim of constraining their astrophysical properties (e.g., age, stellar mass, dust and gas content) and characterizing their role in the context of galaxy evolution. The multi-wavelength coverage of GOODS-S, from X-rays to radio band, allow us to model galaxy SED by using CIGALE with a novel approach, based on a physical motivated modelling of stellar light attenuation by dust. Median stellar mass ($\simeq6.5\times10^{10}$ M$_\odot$) and SFR ($\simeq241$ M$_\odot$ yr$^{-1}$) are consistent with galaxy main-sequence at $z\sim2$. The galaxies are experiencing an intense and dusty burst of star formation (median L$_{\rm IR}\simeq2\times10^{12}$ L$_\odot$), with a median age of $750$ Myr. The high median content of interstellar dust (M$_{\rm dust}\simeq5\times10^8$ M$_\odot$) suggests a rapid enrichment of the ISM (on timescales $\sim10^8$ yr). We derived galaxy total and molecular gas content from CO spectroscopy and/or Rayleigh-Jeans dust continuum ($10^{10}\lesssim$ M$_{\rm gas}/$M$_\odot\lesssim10^{11}$), depleted over a typical timescale $\tau_{\rm depl}\sim200$ Myr. X-ray and radio luminosities (L$_X=10^{42}-10^{44}$ erg s$^{-1}$, L$_{1.5\,{\rm GHz}}=10^{30}-10^{31}$ erg s$^{-1}$, L$_{6\,{\rm GHz}}=10^{29}-10^{30}$ erg s$^{-1}$) suggest that most of the galaxies hosts an accreting radio silent/quiet SMBH. This evidence, along with their compact multi-wavelength sizes 
(median r$_{\rm ALMA}\sim$ r$_{\rm VLA}=1.8$ kpc, r$_{\rm HST}=2.3$ kpc) measured from high-resolution imaging ($\theta_{\rm res}\lesssim$ 1 arcsec), indicates these objects  as the high-z star-forming counterparts of massive quiescent galaxies, as predicted e.g., by the in-situ scenario. Four objects show some signatures of a forthcoming/ongoing AGN feedback, that is thought to trigger the morphological transition from star-forming disks to ETGs.
\end{abstract}



 
\section{Introduction}\label{sec_intro}

Dusty Star-Forming Galaxies (DSFGs) have been discovered almost 20 years ago as an abundant population of distant galaxies characterized by a 850$\mu$m flux density greater than a few mJy \citep[e.g.,][]{Blain2002:2002PhR...369..111B, Hodge2013:2013ApJ...768...91H, CaseyNarayananCooray2014:2014PhR...541...45C, Simpson2015:2015ApJ...807..128S, daCunha2015:2015ApJ...806..110D, Miettinen2015:2015A&A...577A..29M, Oteo2016:2016ApJ...822...36O}.
The subsequent multi-wavelength campaigns \citep[e.g., Hubble Ultra Deep Field, HUDF,][]{Beckwith2006:2006AJ....132.1729B}, along with deep and large-area blind surveys in the infrared domain, e.g. PACS Evolutionary Probe, PEP \citep[][]{Lutz2011:2011A&A...532A..90L}, \textit{Herschel} Multi-tired Extra-galactic Survey, HerMES \citep[][]{Oliver2012:2012MNRAS.424.1614O}, Astrophysical Terahertz Large Area Survey, H-ATLAS \citep[][]{Eales2010:2010PASP..122..499E}, revealed their nature as massive and very infrared luminous galaxies (with typical stellar mass M$_\star\sim$ a few $10^{10}-10^{11}$ M$_\odot$ and infrared luminosity L$_{\rm IR}\sim 10^{11}-10^{12}$ L$_\odot$), characterized by extreme Star Formation Rates  \citep[SFRs $>100$ M$_\odot$ yr$^{-1}$; e.g.][]{Gruppioni2013:2013MNRAS.432...23G, Bethermin2017:2017A&A...607A..89B}.
Due to the intense star formation and the rapid dust enrichment of the Interstellar Medium (ISM), these objects are usually very faint or invisible in the UV/optical domains (even where maps to very low flux density level are available), since dust obscuration is very efficient in these spectral regimes \citep[e.g.][]{Smail1999:1999ApJ...525..609S, Walter2012:2012Natur.486..233W, Franco2018:2018A&A...620A.152F, Williams2019:2019ApJ...884..154W}. In the millimetre band, the cosmological dimming affecting high-z sources is actually offset for these dusty objects by the shifting of the dust peak into the observing band (the so-called \textit{negative k-correction}). As a result, the flux density remains almost constant over a large redshift range, i.e. z $\sim1-10$.

The first spectroscopic follow-up campaigns revealed their number density to peak at redshift $\sim2.5$ \citep[][]{Chapman2003:2003ApJ...599...92C, Chapman2005:2005ApJ...622..772C}, that is almost coincident with the peak of Cosmic Star Formation History \citep[Cosmic SFH; see the review by][]{MadauDickinson2014:2014ARA&A..52..415M} and Black Hole Accretion History \citep[BHAH; e.g.][]{Shankar2009:2009ApJ...690...20S, Aird2010:2010MNRAS.401.2531A, Delvecchio2014:2014MNRAS.439.2736D}. 
On the one hand, DSFGs contribute significantly to both the Cosmic Star Formation Rate Density (Cosmic SFRD) and the stellar mass density, supplying respectively with the $\sim20\%$ and the $\sim 30-50\%$ at $z=2-4$ \citep[see e.g.,][]{Michalowski2010:2010A&A...514A..67M}. 
On the other hand, they possibly assume a central role in the evolution of Active Galactic Nuclei (AGN) and ultimately in the building up of compact quiescient galaxies and massive, local Early Type Galaxies \citep[ETGs; see e.g.,][]{Swinbank2006:2006MNRAS.371..465S, Cimatti2008:2008A&A...482...21C, vanDokkum2008:2008ApJ...677L...5V, Barro2013:2013ApJ...765..104B, Simpson2014:2014ApJ...788..125S, Scoville2017:2017ApJ...837..150S, Toft2014:2014ApJ...782...68T, Oteo2017:2017arXiv170904191O}.

For these reasons, high-z DSFGs represent a crucial tool to solve the complex puzzles of stellar mass assembly and massive galaxies evolution out to $z>3$ \citep[e.g.,][]{Gruppioni2020:2020A&A...643A...8G,Talia2021:2020arXiv201103051T}. Although many progresses have been made in the recent few years, in particular with the advent of ALMA \citep[see the review by][]{ HodgeDaCunha2020:2020arXiv200400934H}, we are far away from fully physically characterizing this class of galaxies and from thoroughly understanding their role in the framework of galaxy formation and evolution.

In order to address these issues, two different (and complementary) approaches are currently on the market, both exploiting the wealth of data recently collected by the numerous wide-area and deep multi-band surveys. 

The first approach aims at constraining one (or a few) specific property of the galaxy population (e.g., stellar mass, dust mass, attenuation, metallicity, environment), relying on the analysis of statistically significant samples of DSFGs \citep[see e.g.,][]{Bethermin2014:2014A&A...567A.103B, daCunha2015:2015ApJ...806..110D, Casey2018:2018ApJ...862...77C, Pearson2018:2018A&A...615A.146P, Franco2018:2018A&A...620A.152F, Donevski2018:2018A&A...614A..33D}. The analysis is often built on a few well-sampled spectral bands \citep[e.g.][]{Magdis2012;2012ApJ...760....6M, Malek2018:2018A&A...620A..50M}, while just in some cases it is multi-messenger \citep[e.g.][]{Pearson2018:2018A&A...615A.146P, Buat2019:2019A&A...632A..79B, Donevski2020:2020A&A...644A.144D}. To preserve the statistical relevance of the outcomes, this approach extensively exploits e.g. the building of synthesized Spectral Energy Distributions (SEDs), especially when the sampling is not uniform for every objects \citep[e.g.][]{Bianchini2019:2019ApJ...871..136B}, and the analysis of stacked data \citep[e.g.][]{Santini2014:2014A&A...562A..30S, Scoville2016:2016ApJ...820...83S}.
Indeed, one of the main drawbacks of this approach is the incompleteness of the data: the available information (e.g., observed frequency, sensitivity, redshift) is typically not homogeneous over the whole sample. In addition, the outcomes may suffer from some selection-bias (depending on the survey exploited to collect the data) and could be affected by uncertainties due to the exploitation of photometric redshift and, eventually, by assumptions on galaxy under-sampled properties.

The second approach focuses on the analysis of small samples or individual objects with a huge number of quality data (both photometric and spectroscopic), with the aim of gaining a deep insight on the ongoing astrophysical processes.
Typically, this objective is reached by exploiting high resolution imaging (usually in the millimetre and radio regimes) and spectral line analysis, that provide an almost secure determination of galaxy spectroscopic redshift, kinematics and size \citep[e.g.,][]{Tadaki2015:2015ApJ...811L...3T, Decarli2016II:2016ApJ...833...70D, Barro2016I:2016ApJ...820..120B, Barro2016II:2016ApJ...827L..32B, Talia2018:2018MNRAS.476.3956T}. At high-z, this analysis requires very long integration times (i.e., orders of a few hours) that do not allow the method to be applied to statistical samples. Some studies partially overcome this problem by focusing on gravitationally lensed objects \citep[e.g.,][]{Negrello2014:2014MNRAS.440.1999N, Massardi2018:2018A&A...610A..53M, Stacey2020:2020MNRAS.tmp.3229S}, even if this strategy requires to model the foreground lens in order to obtain the \textit{original} (i.e., unbent) image of the target galaxy. Although the outcomes do not have any statistical relevance for the whole population of DSFGs, their accuracy can provide an exquisite characterization of the individual object, that encompasses all the galaxy properties, when complemented with the wealth of multi-wavelength photometry currently available \citep[see e.g.,][]{Rujopakarn2016:2016ApJ...833...12R, Elbaz2018:2018A&A...616A.110E}.

In this work we follow the latter approach focusing on a sample of 11 spectroscopically-confirmed z $\sim2$ DSFGs, that we selected in the Great Observatories Origins Survey South \citep[GOODS-S;][]{Dickinson2001:2001AAS...198.2501D, Giavalisco2004:2004ApJ...600L..93G} field, in order to have the widest multi-wavelength coverage of their spectral/broad-band emission currently achievable (from X-rays to radio band). Such accurate sampling is fundamental to derive galaxy astrophysical properties, since their SED is the outcome of all the complex processes occurring between the diverse baryonic components, such as stars and their remnants, cold and warm gas, interstellar dust and central SMBH. We complement the photometric data with high-resolution imaging ($\theta_{\rm res}\lesssim$ 1 arcsec), providing information on galaxy morphology and multi-band sizes, that have been recognized to have a crucial role in studying galaxy formation and evolution \citep[see e.g.,][]{Lapi2018a:2018ApJ...857...22L}. We stress that such a panchromatic approach, combining the outcomes from SED analysis with galaxy spectroscopy and high-resolution imaging, is essential to unbiasedly extract information from multi-band data and shed light on galaxy evolution.

Our main goal is to provide a pilot work on a spectroscopically-confirmed sample of DSFGs at the peak of Cosmic SFH that presents a detailed analysis of the interplay between the ongoing astrophysical processes (such as gas condensation, star formation, central BH accretion and feedback) and a consistent interpretative picture in the framework of galaxy evolution, by exploiting the multiple information coming from photometry, spectroscopy and imaging at high-resolution. Such a novel approach, that easily applies to our small sample of galaxies, could be extended to other multi-band fields \citep[e.g., COSMOS and H-ATLAS; see e.g.][]{Neri2020:2020A&A...635A...7N} and exploited for studying future big samples of spectroscopically-confirmed DSFGs with multi-band follow up (see e.g. the ongoing z-GAL NOEMA Large Program\footnote{http://www.iram.fr/~z-gal/Home.htm\label{footnote1}}; PIs: P. Cox; T. Bakx; H.Dannerbauer). 

This article is organised as follows. In Section~\ref{sec_sample} we describe the selection criteria we use to build our sample of 11 DSFGs and list the multi-wavelength data available for each source; in Section~\ref{sec_fit_with_cigale} we describe the method we follow to model galaxy SEDs, while in Section~\ref{sec_add_info} we illustrate the corresponding outcomes and we match the results with other evidences coming from the available spectral lines and high-resolution imaging; in Section~\ref{sec_results_discussion} we discuss our results by referring to the \textit{in-situ} BH-galaxy co-evolution scenario \citep[see][]{Lapi2018a:2018ApJ...857...22L, Mancuso2016b:2016ApJ...833..152M, Mancuso2017:2017ApJ...842...95M, Pantoni2019:2019ApJ...880..129P}, compare our findings with other recent studies on high-z DSFGs. In Section~\ref{sec_sum_conclusions} we summarize the main outcomes of the present work and we outline our conclusions.

Throughout this work, we adopt the standard flat $\Lambda$CDM cosmology \citep[][]{Planck2018:2020A&A...641A...6P} with rounded parameter values: matter density $\Omega_{\rm M} = 0.32$, dark energy density $\Omega_{\rm \Lambda} = 0.63$, baryon density $\Omega_{\rm b} = 0.05$, Hubble constant $H_0$ = 100 $h$ km s$^{-1}$ Mpc$^{-1}$ with $h$ = 0.67, and mass variance $\sigma_8$ = 0.81 on a scale of 8 $h^{-1}$ Mpc.

\section{The sample}\label{sec_sample}

In order to reach our basic goal, i.e. the accurate sampling of galaxy SED from the X-ray to the radio band, we selected our sample of high-z DSFGs in the (sub-)millimetre regime requiring the following criteria to be fulfilled for each galaxy: 3 or more detections in the optical domain ($\lambda_{\rm obs}=0.3-1$ $\mu$m); 6 or more detections in the NIR+MIR bands ($\lambda_{\rm obs}=1-25$ $\mu$m); 2 or more detections in the FIR band ($\lambda_{\rm obs}=25-400$ $\mu$m);
spectroscopically confirmed redshift in the range $1.5<z<3$; 1 or more detections and/or upper limits in the radio and X-ray regimes.

In more details, we selected our sample from the catalogs by \citet{Yun2012:2012MNRAS.420..957Y}, \citet{Targett2013:2013MNRAS.432.2012T} and \citet{Dunlop2017:2017MNRAS.466..861D}, that are built on (sub-)millimetre surveys of the GOODS-S field that used either ALMA ($\lambda_{\rm B6}= 1.1-1.4$ mm; covering $\sim1$ arcsec$^2$ in the HUDF South), LABOCA ($\lambda=$ 0.870 mm) on APEX or AzTEC ($\lambda=$ 1.1 mm) on ASTE (now on LMT). Then, we exploited the wide and deep broad-band coverage of GOODS-S field to associate the (sub-)millimetre sources with the UV/optical and Infrared photometry currently available in literature and radio and X-ray public catalogs . In particular, we used data from: the MUltiwavelength Southern Infrared Catalog\footnote{GOODS-MUSIC is a multi-band catalog ($\lambda= 0.3 - 8.0\,\mu$m) of NIR (i.e. Z and K$_s$) selected objects in the GOODS-S field. It includes: two images in the U filter by ESO; one image in the B band by the VIMOS/VLT; the ACS/HST images in B, V, i, z bands; the ISAAC-VLT photometry (J, H, K$_s$ bands); the IRAC/\textit{Spitzer} photometry at $\lambda=3.5,\, 4.5,\, 5.8,\, 8\, \mu$m.} \citep[GOODS-MUSIC;][]{Grazian2006:2006A&A...449..951G}, whose UV/optical/NIR photometry was associated with \textit{Spitzer} and \textit{Herschel} infrared photometry by \citet{Magnelli2013:2013A&A...553A.132M}; the $\simeq7$ Ms \textit{Chandra} Deep Field South Catalog \citep[][]{Luo2017:2017ApJS..228....2L}, for the emission in the X-ray; the VLA radio catalogs by \citet{Miller2013:2013ApJS..205...13M}, \citet{Thomson2014:2014MNRAS.442..577T} and \citet{Rujopakarn2016:2016ApJ...833...12R}. 

\textit{Herschel} photometry is crucial to sample the dust FIR peak of z $\sim2$ galaxies and to constrain the dust thermal emission and the intrinsic (i.e., unobscured) stellar light. For these reasons, we included in our study only sources that are detected at least in two \textit{Herschel} (either PACS or SPIRE) photometric bands. Thus, we associated the NIR-selected sources of GOODS-MUSIC catalog with the \textit{Herschel}-detected source in the catalog by \citet{Magnelli2013:2013A&A...553A.132M}. As a result, we obtained a UV-optical-IR photometric catalog made of 263 sources, that we matched with the aforementioned (sub-)millimetre catalogs as we describe in the following.
We associated the millimetre sources observed in the ALMA survey by \citet{Dunlop2017:2017MNRAS.466..861D} using a searching radius of 1 arcsec, compatible with ALMA and HST beam size; the (sub-)millimetre sources by \citet{Yun2012:2012MNRAS.420..957Y} and \citet{Targett2013:2013MNRAS.432.2012T} surveys 
using a 1.5 arcsec searching radius from the optical CANDELS coordinates and a 3 arcsec radius from the NIR IRAC ones (both consistent with the respective beam size). From these two matches we obtained 29 secure associations, 11 in \citet{Dunlop2017:2017MNRAS.466..861D} and 18 in \citet{Yun2012:2012MNRAS.420..957Y} and \citet{Targett2013:2013MNRAS.432.2012T} catalogs.
Due to our selection criteria, these sources are faint, but not completely obscured, in the UV/optical rest-frame band \citep[sampled by ACS/HST and WFC3/HST, with a magnitude limit m$_{\rm AB,\, lim}= 25-26$ mag;][]{Windhorst2010:2010AIPC.1294..225W}, bright in the IR rest-frame domain \citep[sampled by SPIRE/Herschel, with a 5$\sigma$ confusion limited fluxes of $\sim$ 24.0, 27.5, 30.5 mJy at $\lambda=$ 250, 350, 500 $\mu$m, respectively;][]{Nguyen2010:2010A&A...518L...5N,Oliver2012:2012MNRAS.424.1614O}, and detected in the FIR rest-frame with the following flux limits: 1.2 mJy/beam at $\lambda=0.87\,\mu$m (5$\sigma$, including confusion noise; LABOCA); $0.48-0.73$ mJy/beam at $\lambda=1.1$ mm (5$\sigma$, including confusion noise; AzTEC); $35\,\mu$Jy at $\lambda=1.3$ mm (rms; ALMA).

To fulfill our selection criterion on source redshift, we used the ESO compilation of GOODS/CDF-S spectroscopy\footnote{\textit{Spectroscopy Master Catalog}, publicly available at the web page https://www.eso.org/sci/activities/garching/projects/goods.html (updated to 2012).}, the millimetre spectroscopy by \citet{Tacconi2018:2018ApJ...853..179T}, the millimetre catalogs cited above and references therein, and we selected, between the resulting 29 sources, only the ones with spectroscopic redshift in the range $1.5 < z_{\rm spec} < 3$. The latter is a stringent condition in order to properly constrain the physical properties of galaxies by fitting their SED. In particular, we choose all the sources having a measurement of their redshift from optical/millimetre spectral lines with a precision on the third decimal place (at least). The resulting 11 sources are listed in Tab. \ref{tab_redshifts_coords}: the first 7 have an ALMA counterpart, the last 4 an AzTEC-LABOCA.
Hence, we built a multi-wavelength sample composed by 11 (sub-)millimetre selected DSFGs, with a robust spectroscopic measurement of their redshift ($z_{\rm spec}\sim2$).

To complete the multi-band information, we looked for their X-ray counterparts in the $\simeq7$ Ms \textit{Chandra} Deep Field South Catalog by \citet{Luo2017:2017ApJS..228....2L}\footnote{They mapped the CDF-S ($\sim 500$ arcmin$^2$, centered in the GOODS-S field) in the 0.5-7 keV band with an on-axis exposure time of $\simeq 7$ Ms, reaching a sensitivity of $\simeq 1.9\times10^{-17}$ erg s$^{-1}$ cm$^{-2}$. Images in the sub-bands at 0.5-2 keV and 2-7 keV have been then produced and they reach a sensitivity of $\simeq6.4\times10^{-18}$ erg s$^{-1}$ cm$^{-2}$ and $\simeq2.7\times10^{-17}$ erg s$^{-1}$ cm$^{-2}$, respectively.}. We performed a sky match using a search radius of 1.5 arcsec. For nine sources out of the eleven in our sample we found a robust association (separation $\lesssim0.7$ arcsec). No association has been found for the remaining 2 sources, neither in the main or in the ancillary low significance catalogs.

In the radio regime, the 7 (sub-)millimetre sources by \citet{Dunlop2017:2017MNRAS.466..861D} have been followed up at 6 GHz by \citet{Rujopakarn2016:2016ApJ...833...12R} and just 6 are detected. 
\citet{Yun2012:2012MNRAS.420..957Y}, \citet{Targett2013:2013MNRAS.432.2012T} and \citet{Thomson2014:2014MNRAS.442..577T} identified their AzTEC-LABOCA sources in the 1.4 GHz VLA deep map of GOODS-S field by \citet{Miller2013:2013ApJS..205...13M}. From this association all the 4 AzTEC-LABOCA sources in our sample (ALESS067.1, AzTEC.GS25, AzTEC.GS21, AzTEC.GS22) own a robust radio counterpart. Moreover, ALESS067.1 has an additional detection at 610 MHz by the Giant Metre-Wave Radio Telescope \citep[GMRT;][]{Thomson2014:2014MNRAS.442..577T}. As to UDF10, the (sub-)millimetre source without a secure radio detection (i.e., radio flux $<3\sigma$), we take into consideration just the upper limits reported in the aforementioned reference articles.

For completeness, we searched for other ALMA counterparts in the recent catalogs by \citet{Hodge2013:2013ApJ...768...91H} at $\lambda=870\,\mu$m (ALMA B7); \citet{Fujimoto2017:2017ApJ...850...83F}, the so-called DANCING ALMA catalog at $\lambda_{\rm B7}/\lambda_{\rm B6}=0.8-1.1/1.1-1.4$ mm; \citet{Hatsukade2018:2018PASJ...70..105H} at $\lambda_{\rm B6}=1.1-1.4$ mm; \citet{Franco2018:2018A&A...620A.152F} in the ALMA Band 6 and centered at $\lambda=1.13$ mm; and \citet{Cowie2018:2018ApJ...865..106C} in the ALMA Band 7, centered at $\lambda=850\,\mu$m. Three objects selected from LABOCA-AzTEC surveys show robust associations within 1 arcsec (ALESS067.1, AzTEC.GS25, AzTEC.GS21; see Tab. \ref{tab_redshifts_coords}), as well as two objects selected from the ALMA survey by \citet{Dunlop2017:2017MNRAS.466..861D}, i.e. UDF1 and UDF3 (see Tab. \ref{tab_redshifts_coords}). Due to blending, we discarded the AzTEC data for AzTEC.GS21, corresponding to two ALMA detections, and we considered only the ALMA counterpart associated to the IRAC source \citep[i.e., ASAGAO.ID6; see][their Sect. 3.4]{Hatsukade2018:2018PASJ...70..105H}. Finally, we complemented the millimetre continuum with public ALMA data from the ALMA Science Archive\footnote{https://almascience.eso.org/asax/} that will be presented in Pantoni et al. (in preparation), along with the CO line detections that we found for four objects of the sample (UDF1, UDF3, UDF8, ALESS067.1).
In Tab. \ref{tab_multil_photometry} we provide a schematic summary of the currently available photometry (and CO molecule spectroscopy) for each source of the sample, that will be used to perform the SED fitting (Sect.~\ref{sec_fit_with_cigale}) and exploited for the subsequent analysis (Sects. \ref{sec_add_info} and \ref{sec_results_discussion}).

We note that most of the studies on high-z DSFGs conduced in the last couple of years by exploiting SED fitting \citep[e.g., ][]{daCunha2015:2015ApJ...806..110D, Casey2017:2017ApJ...840..101C, Franco2020:2020A&A...643A..30F, Donevski2020:2020A&A...644A.144D, Dudzeviciute2021:2021MNRAS.500..942D} focuses on, e.g., galaxy location with respect the star-forming main-sequence \citep[i.e. the empirical relation between stellar mass and SFR followed by star-forming galaxies, see][]{Daddi2007:2007ApJ...670..156D, Noeske2007:2007ApJ...660L..43N}; the search of diagnostic quantities, such as the dust-to-stellar mass ratio, in order to disentangle high-z main-sequence and starburst galaxies and probing the evolutionary phase of massive objects; the evolution of SFR, stellar mass, stellar attenuation and dust mass with redshift; the difference between populations of DSFGs selected in the FIR or in the (sub-)millimetre domain; the link between star formation surface density and gas depletion time, and their role in determining the galaxy subsequent evolution. 

In this work, on the one hand, we reduce the uncertainties on the constrained parameters by both selecting solely the sources with spectroscopically confirmed redshift at the peak of Cosmic SFH and requiring an accurately-sampled SED from the X-ray to radio regime, with particular care on the FIR peak and the optical/NIR emission by stars, that are essential to constrain interstellar dust attenuation, galaxy age and SFR. Moreover, when combined together, these requirements allow to get an insight on both the role of DSFGs in the cosmic stellar mass assembly and galaxy co-evolution with central SMBH. 

On the other hand, the requirement for a complete multi-band coverage and the condition on the availability of spectroscopic redshift do not allowed us to include in our sample objects that are totally obscured in the UV/optical, and seriously limit our sample size and completeness. Even if we are not able to 
trace statistical indications, it is enough to test our approach to extract information on the sources, as we will describe in the following Sections.

\begin{table*}
\centering
\caption{The 11 (sub-)millimetre sources of our sample. Identification codes (ID) in the first two columns refer to: *) GOODS (IAU); **) (sub-)millimetre surveys available in the literature \citep[][]{Yun2012:2012MNRAS.420..957Y, Hodge2013:2013ApJ...768...91H, Targett2013:2013MNRAS.432.2012T, Dunlop2017:2017MNRAS.466..861D}. Spectroscopic redshifts and ALMA, HST and LABOCA-AzTEC (labelled as mm) sky positions are listed in the central columns. 
References for ALMA counterparts:  
$\triangle$) \citet{Hodge2013:2013ApJ...768...91H};  
$\spadesuit$) \citet{Fujimoto2017:2017ApJ...850...83F};
$\star$) \citet{Dunlop2017:2017MNRAS.466..861D};
$\diamondsuit$) \citet{Hatsukade2018:2018PASJ...70..105H};
$\circ$) \citet{Cowie2018:2018ApJ...865..106C};
$\bullet$) \citet{Franco2018:2018A&A...620A.152F};
$\heartsuit$) Pantoni et al. (in preparation).
References for spectroscopic redshifts (uncertainty at least on the third decimal digit): 
$a$) \citet{Szokoly2004:2004ApJS..155..271S} found z$_{\rm H_\alpha}=2.688\pm0.005$, consistent with Pantoni et al. (in preparation) measurement within 2$\sigma$; 
$b$) \citet{Momcheva2016:2016ApJS..225...27M}; 
$c$) \citet{Kurk2013:2013A&A...549A..63K}; 
$d$) \citet{Dunlop2017:2017MNRAS.466..861D}; 
$e$) Pantoni et al. (in preparation); 
$f$) \citet{Straughn2009:2009AJ....138.1022S};
$g$) \citet{Popesso2009:2009A&A...494..443P};
$h$) \citet{Vanzella2008:2008A&A...478...83V}; 
$i$) \citet{Targett2013:2013MNRAS.432.2012T} and reference therein;
$j$) \citet{Kriek2006:2006ApJ...649L..71K, Kriek2007:2007ApJ...669..776K} found z$_{\rm H_\alpha}=2.122\pm0.053$, consistent with Pantoni et al. (in preparation) measurement within the uncertainties.
}\label{tab_redshifts_coords}
\resizebox{2.1\columnwidth}{!}{
\hspace{-0.5cm}
\begin{tabular}{lcccccccc}
\hline
\bfseries {ID}* & \bfseries {ID}** & $\mathbf{z_{spec}}$ & $\mathbf{RA_{ALMA}}$ &$\mathbf{DEC_{ALMA}}$ & $\mathbf{RA_{HST}}$
&$\mathbf{DEC_{HST}}$& $\mathbf{RA_{mm}}$ &$\mathbf{DEC_{mm}}$ \\
& &  & [h:m:s] &[\textdegree:\textquotesingle:\textquotesingle\textquotesingle]& [h:m:s] & [\textdegree:\textquotesingle:\textquotesingle\textquotesingle]& [h:m:s] & [\textdegree:\textquotesingle:\textquotesingle\textquotesingle] \\
\hline 
J033244.01-274635.2$^{\star,\,\circ,\,\bullet,\,\heartsuit}$ & UDF1 & $2.698\pm0.002^{(e,a)}$ & 03:32:44.04 & -27:46:36.01 &  03:32:44.04 & -27:46:36.01 & $-$ & $-$ \\
J033238.53-274634.6$^{\star,\,\circ,\,\bullet,\,\heartsuit}$  &UDF3 & $2.543\pm0.005^{(e)}$ & 03:32:38.55 & -27:46:34.57 & 03:32:38.55 & -27:46:34.57 & $-$ & $-$\\
J033236.94-274726.8$^{\star}$ & UDF5 & $1.759\pm0.008^{(b)}$& 03:32:36.96 & -27:47:27.13 & 03:32:36.97 & -27:47:27.28 & $-$ & $-$\\
J033239.74-274611.4$^{\star}$ & UDF8 & $1.549\pm0.005^{(e)}$ &03:32:39.74 & -27:46:11.64 & 03:32:39.73 & -27:46:11.24 & $-$ & $-$ \\
J033240.73-274749.4$^{\star}$ & UDF10 & $2.086\pm0.006^{(b)}$ & 03:32:40.75 & -27:47:49.09 & 03:32:40.73 & -27:47:49.27 & $-$ & $-$ \\
J033240.06-274755.5$^{\star}$ & UDF11 & $1.9962\pm0.0014^{(c,\,d)}$ & 03:32:40.07 & -27:47:55.82 & 03:32:40.06 & -27:47:55.28 & $-$ & $-$\\
J033235.07-274647.6$^{\star}$ & UDF13 & $2.497\pm0.008^{(b)}$ & 03:32:35.09 & -27:46:47.78 & 03:32:35.08 & -27:46:47.57 & $-$ & $-$ \\
J033243.19-275514.3$^{\triangle,\,\heartsuit}$  & ALESS067.1 & ${2.1212^{+0.0014}_{-0.0005}}^{(e,\,j)}$ & 03:32:43.20 & -27:55:14.16&03:32:43.3  & -27:55:17 & 03:32:43.18 & -27:55:14.49 \\ 
J033246.83-275120.9$^{\spadesuit,\,\circ}$  & AzTEC.GS25 & $2.292\pm0.001^{(g)}$ & 03:32:46.83  &  -27:51:20.97 &  03:32:46.96 & -27:51:22.4 & 03:32:46.84 & -27:51:21.23 \\
J033247.59-274452.3$^{\diamondsuit,\,\circ,\,\heartsuit}$  & AzTEC.GS21 & $1.910\pm0.001^{(h)}$ & 03:32:47.59 & -27:44:52.43 & 03:32:47.60 & -27:44:49.3 & 03:32:47.59 & -27:44:52.32 \\
J033212.55-274306.1$^{\heartsuit}$ & AzTEC.GS22 & $1.794\pm0.005^{(i)}$ & 03:32:12.52 & -27:43:06.00 &03:32:12.60 & -27:42:57.9 & 03:32:12.56 & -27:43:05.37\\
\hline
\end{tabular}
}
\end{table*}

\begin{table*}
\centering
\caption{Up-to-date available UV-optical, infrared, (sub-)millimetre, radio and X-ray photometry and other spectral information (CO line detection) for our sample. UV-optical and NIR photometry from MUSIC catalog \citep[][]{Grazian2006:2006A&A...449..951G}:
$a$) ACS-HST, $\lambda_{\rm c} = 0.433,\,0.594,\,0.771,\,0.886\,\mu m$ ;
$b$) ISAAC-VLT, $\lambda_{\rm c} = 1.255,\,1.656,\,2.163\,\mu m$ ;
$c$) Spitzer, $\lambda_{\rm c} = 3.6,\,4.5,\,5.8,\,8,\,16,\,24\,\mu m$. 
FIR photometry from \citet{Magnelli2011:2011A&A...528A..35M,Magnelli2013:2013A&A...553A.132M}, as result from the sky-match between GOODS-MUSIC catalog and MIPS-Herschel photometry:
$d$) PACS-Herschel, $\lambda_{\rm c} = 70,\,100,\,160 \,\mu m$ ;
$e$) SPIRE-Herschel, $\lambda_{\rm c} = 250,\,350,\,500\,\mu m$. 
millimetre and radio photometry from \citet{Miller2013:2013ApJS..205...13M}, \citet{Yun2012:2012MNRAS.420..957Y}, \citet{Hodge2013:2013ApJ...768...91H}, \citet{Targett2013:2013MNRAS.432.2012T}, \citet{Thomson2014:2014MNRAS.442..577T}, \citet{Dunlop2017:2017MNRAS.466..861D}, \citet{Rujopakarn2016:2016ApJ...833...12R}, \citet{Fujimoto2017:2017ApJ...850...83F}, \citet{Hatsukade2018:2018PASJ...70..105H}, \citet{Franco2018:2018A&A...620A.152F}, \citet{Cowie2018:2018ApJ...865..106C} resulting from the cross-match with GOODS-MUSIC catalog: 
$f$) LABOCA/APEX, $\lambda_{\rm c} = 870\,\mu m$ ;
$g$) ALMA, $\lambda_{\rm c} = 0.450,\,0.870-1,\,1.3,\,2,\,3$ mm;
$h$) AzTEC/ASTE, $\lambda_{\rm c} =1100 \,\mu m$ ;
$i$) JVLA, $\nu_{\rm c} = 6,\, 1.49$ GHz;
$j$) GMRT, $\nu=610$ MHz.
X-ray data resulting from the catalog by \citet{Luo2017:2017ApJS..228....2L}:
$k$) Chandra ACIS, $0.5-7.0$ keV.
CO spectral lines by Pantoni et al. (in preparation).
Checkmark stands for \textit{detection}; nd stands for \textit{non-detection}/\textit{upper-limits}; void means the source has not been observed with the corresponding device. Brackets mean that the source is detected but the data has not be taken in consideration.}\label{tab_multil_photometry}
\begin{tabular}{lcccccccccccccccc}
\hline
\bfseries {ID}  & $\mathbf{B}^a$& $\mathbf{V}^a$& $\mathbf{i}^a$& $\mathbf{z}^a$&
$\mathbf{J}^b$& $\mathbf{H}^b$& $\mathbf{Ks}^b$&$\mathbf{IR_{36}}^c$&$\mathbf{IR_{45}}^c$&$\mathbf{IR_{58}}^c$& $\mathbf{IR_{80}}^c$&
$\mathbf{IRS_{16}}^c$&$\mathbf{F_{24}}^c$&$\mathbf{F_{70}}^d$&$\mathbf{F_{100}}^d$& $\mathbf{F_{160}}^d$\\
\hline
UDF1 &\checkmark& \checkmark&  \checkmark&\checkmark & \checkmark& \checkmark&\checkmark& \checkmark&
\checkmark & \checkmark& \checkmark& nd &\checkmark& nd &\checkmark &\checkmark \\
UDF3 &\checkmark& \checkmark&\checkmark&\checkmark & \checkmark& \checkmark&\checkmark& \checkmark& \checkmark & \checkmark& \checkmark&\checkmark& \checkmark& nd & \checkmark &\checkmark\\
UDF5  & \checkmark& \checkmark& \checkmark&\checkmark & \checkmark& \checkmark&\checkmark& \checkmark& \checkmark & \checkmark& \checkmark&\checkmark& \checkmark& nd &\checkmark &\checkmark\\
UDF8 & nd &\checkmark&\checkmark&\checkmark & \checkmark& \checkmark&\checkmark& \checkmark& \checkmark & \checkmark& \checkmark&\checkmark& \checkmark& nd & \checkmark &\checkmark\\
UDF10  & \checkmark& \checkmark& \checkmark&\checkmark & \checkmark& \checkmark&\checkmark& \checkmark&\checkmark & \checkmark& \checkmark& nd & nd & nd &\checkmark &\checkmark\\
UDF11  & \checkmark& \checkmark&\checkmark&\checkmark & \checkmark& \checkmark&\checkmark& \checkmark&\checkmark & \checkmark& \checkmark&\checkmark&\checkmark& nd &\checkmark &\checkmark\\
UDF13 & \checkmark&  \checkmark& \checkmark&\checkmark & \checkmark& \checkmark&\checkmark& \checkmark& \checkmark & \checkmark& \checkmark& nd &\checkmark& nd &\checkmark & nd\\
ALESS067.1  & \checkmark& \checkmark& \checkmark&\checkmark & \checkmark& \checkmark&\checkmark& \checkmark& \checkmark & \checkmark& \checkmark& nd &\checkmark& nd &\checkmark &\checkmark\\
AzTEC.GS25  & \checkmark& \checkmark& \checkmark&\checkmark & \checkmark& nd &\checkmark& \checkmark& \checkmark & \checkmark& \checkmark& nd &\checkmark& nd &\checkmark &\checkmark\\
AzTEC.GS21  & \checkmark& \checkmark& \checkmark&\checkmark & \checkmark& nd &\checkmark& \checkmark& \checkmark & \checkmark& \checkmark& \checkmark& \checkmark&\checkmark &\checkmark &\checkmark\\
AzTEC.GS22 & nd & \checkmark& \checkmark&\checkmark & \checkmark& \checkmark&\checkmark& \checkmark& \checkmark & \checkmark& \checkmark& nd &\checkmark& nd & nd &\checkmark \\
\hline 
\end{tabular}

\begin{tabular}{ccccccccccccccc}
\hline
$\mathbf{F_{250}}^e$&$\mathbf{F_{350}}^e$&$\mathbf{B9}^g$&$\mathbf{F_{500}}^e$
&$\mathbf{F_{870}}^f$ &$\mathbf{B7}^g$ &$\mathbf{F_{1100}}^h$&$\mathbf{B6}^g$&$\mathbf{B4}^g$&$\mathbf{B3}^g$ &$\mathbf{C\,band}^i$&$\mathbf{L\,band}^i$ &$\mathbf{610 MHz}^j$ &$\mathbf{X-ray}^k$& 
$\mathbf{CO \,line}$\\
\hline
 \checkmark& \checkmark& &\checkmark & & \checkmark& & \checkmark& & & \checkmark& & & \checkmark& J(3-2)\\
nd & nd & & nd & & \checkmark & & \checkmark& & &\checkmark& & &\checkmark&J(3-2)\\
\checkmark&\checkmark& & nd & & & & \checkmark& & & \checkmark& & & nd &\\
\checkmark&\checkmark& & nd & & & & \checkmark& & & \checkmark& & &\checkmark&J(2-1)\\
nd & nd & & nd & & & & \checkmark& & & nd& & &\checkmark&\\
\checkmark&\checkmark& & nd & & & & \checkmark& & & \checkmark& & &\checkmark&\\
nd &\checkmark& & nd & & & & \checkmark& & &  \checkmark& & &\checkmark&\\
\checkmark&\checkmark& & \checkmark&\checkmark & \checkmark &nd & & \checkmark& \checkmark & & \checkmark& \checkmark& \checkmark&J(3-2)\\
\checkmark&\checkmark& & \checkmark& &\checkmark & \checkmark& & & &  & \checkmark& & \checkmark&\\
\checkmark&\checkmark& \checkmark&nd & &\checkmark & (\checkmark) & \checkmark& & &  & \checkmark& &\checkmark&\\
\checkmark&\checkmark &\checkmark& nd & & & \checkmark & & & &  & \checkmark& & nd &\\
 \hline
\end{tabular}
\end{table*}

\section{SED fitting with CIGALE}\label{sec_fit_with_cigale}

We use the Code Investigating GAlaxy Emission \citep[CIGALE;][]{Boquien2019:2019A&A...622A.103B} to model the SEDs of our 11 high-z DSFGs. CIGALE is a Python code that can reproduce galaxy broad-band emission - from FUV to radio wavelengths - by physically preserving an energy balance between stellar light emitted in optical/UV and re-radiated in IR. To this aim, a large grid of models is fitted to the data. 
In a nutshell, CIGALE builds composite stellar populations by combining single stellar population with flexible SFHs, calculates the ionized gas emission from young stars and exploits flexible attenuation curves to attenuate both the stellar and the ionized gas emission. Infrared emission is estimated by balancing the energy absorbed from the optical/UV domain and the energy re-emitted at longer wavelengths by dust. 
The main physical properties, such as SFR, attenuation, dust luminosity, stellar mass, AGN fraction, are estimated  by means of $\chi^2$ and bayesian analysis.
In this Section we briefly describe the modules we have exploited to model galaxy SEDs and explain which are the main motivations that have driven our choices.
In Tab. \ref{tab_cigale_param_priors} we provide the complete list of the modules we used and the priors we assumed for the corresponding free parameters.

\begin{table*}
\centering
\caption{Input parameters configuration used to fit the broad-band emission of our z $\sim2$ DSFGs with CIGALE. Columns show (in order): the input CIGALE modules we used (\textbf{CIGALE module}); the functional shape assumed for the modelled quantity (when required; \textbf{Shape}); list of module free parameters (units in square brackets; \textbf{Free parameters}); free parameter symbol (\textbf{Symbol}); prior values for the corresponding parameter (\textbf{Values}). References for the adopted shapes are: $a$) \citet{Mancuso2016b:2016ApJ...833..152M}; $b$) \citet{BruzualCharlot2003:2003MNRAS.344.1000B} ; $c$) \citet{LoFaro2017:2017MNRAS.472.1372L} ; $d$) \citet{Casey2012:2012MNRAS.425.3094C}. }\label{tab_cigale_param_priors}
 \begin{tabular}{ccccc}
 \hline
 \bfseries {CIGALE module} & \bfseries {Shape} & \bfseries {Free parameters} & \bfseries {Symbol} & \bfseries {Values} \\
 \hline
 Star Formation History  & Constant$^a$ &  Age [Gyr] & $\tau_{\star}$ & 2, 1.5, 1.0, 0.8, 0.7, 0.6, 0.5, 0.4, 0.3, 0.2, 0.15 \\
 \hline
 & &  Initial mass function & IMF & Chabrier (fixed)\\
 Single Stellar Population & Two SSPs$^b$ & Metallicity & $Z$ &  0.02 (fixed)\\
 & & Separation age [Myr] & $\Delta_\star$ & 10, 20, 30, 50, 100, 150, 200\\
 \hline
 &  & V-band attenuation in BCs & $A_{\rm V}^{\rm BC}$ & 10000 (fixed) \\
 Dust Attenuation  & Double power-law$^c$ & Power-law slope in BCs & $\delta_{\rm BC}$ & -1.6, -1.8, -2.0, -2.2, -2.4, -2.6 \\
 & & $A_{\rm V}^{\rm ISM}$/$A_{\rm V}^{\rm BC}$ & $f_{\rm att}$ & 0.0001, 0.0002, 0.0003\\
 & & Power-law slope in ISM & $\delta_{\rm ISM}$ &  -0.7, -0.5, -0.3, -0.1\\
 \hline
 & Power-law +&  MIR power-law slope & $\alpha_{\rm dust}$ & 1.6, 1.8, 2.0, 2.2 \\
 Dust Emission  &  Single-T Modified BB$^d$ & Dust emissivity index & $\beta_{\rm dust}$ & 1.6, 1.8, 2.0, 2.2, 2.4, 2.6 \\
 &  & Dust temperature [K] & T$_{\rm dust}$ & 20, 30, 40, 50, 60, 70 \\
 \hline
\end{tabular}
\end{table*}

\subsection{Star Formation History}

One of the more influential but yet scarcely constrained components of SED fitting is the functional form of galaxy Star Formation History (SFH). Since a large variation in the SFH can yield similar SEDs, broad-band fitting alone can not really provide certain information on the past star formation phases of galaxies. This problem led to the exploitation of quite simple SFHs, such as constant, exponential (decaying or rising), delayed or periodic \citep[e.g.,][]{Elbaz2018:2018A&A...616A.110E}; in the case of high-z DSFGs, additional bursts of star formation are often included \citep[e.g.,][]{Ciesla2017:2017A&A...608A..41C, Forrest2018:2018ApJ...863..131F, Donevski2020:2020A&A...644A.144D}. 

In this work, we assume a delayed exponential SFH, basing our choice on many studies of SED-modelling for high-z star-forming galaxies \citep[e.g.,][]{Papovich2011:2011MNRAS.412.1123P, Smit2012:2012ApJ...756...14S, Moustakas2013:2013ApJ...767...50M, Steinhardt2014:2014ApJ...791L..25S, Cassara2016:2016A&A...593A...9C, Citro2016:2016A&A...592A..19C}, that suggest a slow power-law increase of galaxy SFR over a timescale $\tau_{\star}$ (i.e., burst duration), followed by a rapid exponential decline. In particular, for $\tau\lesssim\tau_{\star}$, we adopt the functional form used by \citet{Mancuso2016b:2016ApJ...833..152M}:
\begin{equation}\label{eq_SFH_Mancuso16b}
    {\rm SFR}(\tau) \propto \biggl(\frac{\tau}{\tau_\star}\biggr)^{0.5} \quad \quad {\rm for}\quad \tau\lesssim\tau_\star
\end{equation}
(see their Eq. 3 and Fig. 3, top panel, dashed line). We simplify the evolution for $\tau>\tau_{\star}$ by assuming that star-formation stops at $\tau\sim\tau_{\star}$, quenched by AGN feedback. The latter assumption is validated by the observed fraction of far-IR-detected host galaxies in the  X-ray \citep[e.g.,][]{Mullaney2012:2012MNRAS.419...95M, Page2012:2012Natur.485..213P, Rosario2012:2012A&A...545A..45R} and optically selected AGNs \citep[e.g.,][]{Mor2012:2012ApJ...749L..25M, Wang2013:2013ApJ...773...44W, Willott2015:2015ApJ...801..123W}, indicating that SFR in massive ($M_{\star}>10^{10}$ M$_\odot$) galaxies is abruptly stopped by AGN feedback, after $\tau_{\star}$, over a very short timescale, i.e. $<10^7-10^8$ yr.
As such, the adopted SFH has just one free parameter, i.e. the main star-forming burst duration ($\tau_\star$). We assume $\tau_\star$ to vary in the range $[0.15-2]$ Gyr (see table~\ref{tab_cigale_param_priors}), as it is suggested by recent observations with ALMA, characterising the dust-enshrouded star formation of massive high-redshift galaxies \citep[e.g.,][]{Scoville2014:2014ApJ...783...84S, Scoville2016:2016ApJ...820...83S}, and by the observed $\alpha$-enhancement, i.e. iron underabundance compared to $\alpha$ elements, in local ellipticals \citep[see the reviw by][]{Renzini2006:2006ARA&A..44..141R}.

\subsection{Stellar component}\label{sec_stellar_component}

We compute the stellar light following the prescriptions in \citet{BruzualCharlot2003:2003MNRAS.344.1000B}. We divide the stellar population into two Simple Stellar Populations (SSPs): young stars, which are supposed to be completely enshrouded into their molecular Birth Clouds (BCs); old stars, which live in the ISM and have already dissolved their BCs. We set the separation age between these two components as a free parameter, fixing the priors (see Tab. \ref{tab_cigale_param_priors}) to the typical values found to be valid for star-forming progenitors of massive ellipticals \citep[e.g.,][]{Schurer2009:2009MNRAS.394.2001S}. We assume a \citet{Chabrier2003:2003PASP..115..763C} Initial Mass Function (IMF; $M_\star = 0.1-100$ M$_\odot$) and we set the stellar metallicity to the solar one, i.e. $Z=0.02$.

\subsection{Stellar light attenuation by dust}\label{sec_dust_att}

In order to model interstellar dust attenuation in the UV/optical, we adopt a double power-law \citep[][see their Eqs. 5 and 6]{LoFaro2017:2017MNRAS.472.1372L}, reproducing the different contributions from young and old stars.

We prescribe the light  from young stars to be completely absorbed by dust in the surrounding BC by fixing the BC attenuation in the V-band to the value $A_{\rm V}^{\rm BC} =10^4$. The energy balance provided by CIGALE ensures that the light absorbed by dust is re-emitted in the FIR (i.e., dust thermal emission). To be self-consistent, we fix the slope $\delta_{\rm BC}$ to the dust FIR spectral index ($\beta_{\rm dust}$) best value, as it is obtained by the fitting procedure.

After $t_0 \sim 10^7$ yr from their birth, old stars have already dissipated/escaped their BC, and the light they emit is expected to be attenuated just by dust populating galaxy ISM (whose V-band attenuation is defined by the relation $f_{att}=A_{\rm V}^{\rm ISM}/A_{\rm V}^{\rm BC}$). Following the prescriptions in \citet{LoFaro2017:2017MNRAS.472.1372L}, we allow the ISM attenuation slope $\delta_{\rm ISM}$ to span the range: [-0.7, -0.5, -0.3, -0.1] (see Tab. \ref{tab_cigale_param_priors}). 

We refer the reader to Appendix \ref{appendix_attlaw} for the detailed derivation of individual galaxy attenuation law (see Fig. \ref{panel}, right column).

\subsection{Dust emission}\label{sec_dust_emission}

We model the IR interstellar dust emission by decomposing the IR light into two different components: a power-law, describing the mid-IR (MIR) emission coming from Polycyclic Aromatic Hydrocarbon (PAHs) and central AGN; a single-temperature modified black-body (BB), describing the far-IR (FIR) thermal emission of cold interstellar dust, as suggested by \citet{Casey2012:2012MNRAS.425.3094C}. This simple approach is very convenient in our case since the MIR SED of our 11 DSFGs is scarcely sampled (each source owns just one or two photometric data points in this spectral range). We exploit the fitting formula shown in \citet[her Eq. 3]{Casey2012:2012MNRAS.425.3094C}, providing a more accurate fit to those systems with fewer MIR photometric data. The function has three free parameters: the MIR power-law slope $\alpha_{\rm dust}$, the dust emissivity index $\beta_{\rm dust}$ and the dust temperature T$_{\rm dust}$.

We set the spectral index $\alpha_{\rm dust}$ to vary in the range [1.6, 1.8, 2.0, 2.2], following the results obtained by \citet{Mullaney2011:2011MNRAS.414.1082M}.
The value of $\beta_{\rm dust}$ is usually assumed to be $1.5$ and historically ranges between $1-2$ \citep[see e.g.,][]{Hildebrand1983:1983QJRAS..24..267H}. Some recent works suggest a wider range for $\beta_{\rm dust}$, between 1 and 2.5 \citep[e.g.][]{Casey2011:2011MNRAS.411.2739C,Chapin2011:2011MNRAS.411..505C, Gilli2014:2014A&A...562A..67G, Bianchi2013:2013A&A...552A..89B, Bianchini2019:2019ApJ...871..136B, Pozzi2020:2020MNRAS.491.5073P}, favouring the higher values in the range (i.e. $\beta_{\rm dust}>1.5$). For this reason, we set the dust spectral index $\beta_{\rm dust}$ to freely vary in the range $[1.6 - 2.6]$. The dust temperature $T_{\rm dust}$ can vary between $[20 - 70]$ K, which almost corresponds to the normal range of dust temperatures expected for galaxy ISM heated only by star formation (see Tab. \ref{tab_cigale_param_priors}). A more realistic approach is the one assuming more than one family of cold dust grains (i.e., with different temperatures) populating the galaxy ISM. However, fitting the dust thermal emission with a multi-temperature modified BB requires at least more than five photometric data points in the FIR regime, which are currently not available for every source of the sample.

We avoid to model the AGN MIR emission alone since the most quoted AGN models \citep[e.g.,][]{Fritz2006:2006MNRAS.366..767F, Nenkova2008:2008ApJ...685..147N, Feltre2012:2012MNRAS.426..120F} are extremely complex and characterized by a number of free parameters that would be hardly constrained by the under-sampled MIR emission of our galaxies. In addition, considering the absence of any spectral information in the MIR band, we fairly avoid the exploitation of more sophisticated models that include PAH and silicate lines, such as \citet{DraineLi2007:2007ApJ...657..810D}. We would like to stress that these choices do not affect significantly the derived value of the IR total luminosity. Indeed, the net effect of the MIR emission coming from both AGN and PAHs on the integrated IR luminosity attains less than 10-20\%, meaning that the cold-dust modified BB still dominates the bulk of the total IR emission, when integrated. This fact has been recently demonstrated by the analysis of a (local) sample of IR galaxies hosting a moderate powerful X-ray AGN (i.e., $L_{\rm 2-10\,keV}\sim10^{42}-10^{44}$ erg s$^{-1}$) by \citet{Mullaney2011:2011MNRAS.414.1082M}, and it is still valid for our 11 DSFGs (as we discuss in Sect.~\ref{sec_Xray_analysis}).

Finally, it is important to take into consideration that the method we use is \textit{luminosity-weighted}: dust temperature is estimated from the modified BB FIR peak, which is more sensitive to the warmest population of dust grains. As such, the single-temperature SED fitting procedure tends to return values of $T_{\rm dust}$ higher than the mean one. This tendency and how it affects the SED-derived dust mass have been widely discussed and investigated in the last decade, both in local and high-z Universe \citep[e.g.,][]{Dale2012:2012ApJ...745...95D, Magdis2012;2012ApJ...760....6M, Berta2016:2016A&A...587A..73B, Schreiber2018:2018A&A...609A..30S, Liang2019;2019MNRAS.489.1397L, Martis2019:2019ApJ...882...65M}. \citet{Magdis2012;2012ApJ...760....6M} quantify this effect on a statistical sample of z $\gtrsim 2$ Sub-Millimetre Galaxies (SMGs) finding that the fit with single-temperature modified BB gives dust masses that are (on average) lower by a factor of $\sim2$ when compared with \citet{DraineLi2007:2007ApJ...657..810D} model. We will extensively comment on this result in Sects. \ref{sec_dust_mass} and \ref{sec_gas_mass}, where we discuss the derivation of dust and gas masses for our sample.

\begin{table*}
\centering
\caption{CIGALE outputs from stellar (star formation $+$ emission $+$ attenuation) bayesian analysis for the sources of the sample (ID and spectroscopic redshift in the first two columns). Third column shows the corresponding best-fit reduced $\chi^2$. In the order, we list the outcomes from: SFH module (SFR and galaxy age $\tau_\star$, i.e. the burst duration), stellar emission (stellar mass, M$_\star$; age separation between old and young stars, $\Delta_\star$; restituted gas mass to ISM from stellar evolution, M$_{\rm R}$) and stellar attenuation (ISM-to-BC V-band attenuation, $f_{\rm att}$; BC attenuation spectral index, $\delta^{\rm BC}$; ISM attenuation spectral index, $\delta^{\rm ISM}$). Units are indicated between square brackets.}\label{tab_results_bayes_stars}
  \begin{tabular}{lcccccccccc}
 \hline
 \bfseries {ID} & $\mathbf{z_{spec}}$ & $\mathbf{ \chi_{red}^2}$ & \bfseries {SFR}& $\mathbf{\tau_\star}$ & $\mathbf{M_\star}$ &$\mathbf{\Delta_\star}$ & $\mathbf{M_{R}}$ & $\mathbf{f_{att}}$ & $\mathbf{\delta^{BC}}$ & $\mathbf{\delta^{ISM}}$ \\
  & & & [M$_\odot$ yr$^{-1}$] & [Myr] & [$10^{10}$ M$_{\odot}$] & [Myr] & [$10^{10}$ M$_{\odot}$] & [$10^{-4}$] &  &  \\
 \hline
  UDF1 & 2.698 & 1.15 & $352\pm18$ & $334\pm58$ & $8\pm1$ & $10.0\pm0.5$ & $3.3\pm0.6$ & $3.0\pm0.2$ & $-2.1\pm0.3$ & $-0.1\pm0.003$ \\
  UDF3  & 2.543 & 3.28  & $519\pm38$ & $234\pm47$ & $9\pm1$ & $30\pm4$ & $3.2\pm0.6$ & $3.0\pm0.2$ & $-2.1\pm0.3$ & $-0.1\pm0.002$ \\
  UDF5 & 1.759 & 1.69 & $85\pm6$ & $404\pm85$ & $2.4\pm0.3$ & $20\pm1$ &  $1.0\pm0.2$ & $1.0\pm0.05$ & $-2.1\pm0.3$  & $-0.7\pm0.003$ \\
  UDF8 & 1.549 & 2.27 & $100\pm5$ & $992\pm50$ & $6.5\pm0.3$ & $160\pm23$ & $3.4\pm0.2$ & $2.0\pm0.1$ & $-2.4\pm0.3$ & $-0.7\pm0.003$ \\
  UDF10 & 2.086 & 1.02 & $41\pm5$ & $917\pm137$ & $2.5\pm0.3$ & $50\pm9$ & $1.2\pm0.2$ & $1.0\pm0.05$ & $-2.1\pm0.3$ & $-0.54\pm0.08$ \\
  UDF11 & 1.9962 & 1.76 & $241\pm19$ & $380\pm82$ &  $6.4\pm0.9$ & $51\pm9$ & $2.6\pm0.5$ & $1.0\pm0.05$ & $-2.1\pm0.3$ & $-0.49\pm0.09$\\
  UDF13 & 2.497& 0.99 & $111\pm17$ & $879\pm149$ & $6.5\pm1.4$ & $51\pm58$ & $3.3\pm0.8$ & $1.8\pm0.4$ & $-2.1\pm0.3$ & $-0.6\pm0.1$  \\
  ALESS067.1 & 2.1212 & 1.73  & $487\pm24$ & $903\pm100$ & $29\pm3$ & $10.0\pm0.6$ & $15\pm2$ & $3.0\pm0.2$ & $-1.8\pm0.3$ & $-0.3\pm0.003$ \\
  AzTEC.GS25 & 2.292 & 1.77 & $401\pm20$ & $290\pm88$ & $8\pm2$ &   $29\pm20$ & $3\pm1$ & $2.5\pm0.5$ & $-2.1\pm0.3$ & $-0.1\pm0.008$  \\
  AzTEC.GS21 & 1.910 & 1.76 & $360\pm18$ & $746\pm105$ & $18\pm2$ &  $50\pm3$ & $9\pm1$ & $3.0\pm0.5$ & $-2.0\pm0.3$  & $-0.5\pm0.004$ \\
  AzTEC.GS22 & 1.794 & 1.37 & $91\pm5$ & $940\pm74$ & $5.7\pm0.5$ & $11\pm4$ & $2.9\pm0.3$ & $3.0\pm0.5$ & $-2.4\pm0.3$ & $-0.7\pm0.01$ \\
  \hline
  \end{tabular}
  \end{table*}
  
 \begin{table}
 \centering
  \caption{CIGALE outputs from dust emission bayesian analysis for the sources of the sample (ID in the first column). In the order, we list: MIR power-law spectral index ($\alpha_{\rm dust}$); FIR modified-BB spectral index ($\beta_{\rm dust}$); dust temperature (T$_{\rm dust}$) and dust luminosity (L$_{\rm dust}$) in units of $10^{12}$ L$_{\odot}$.}\label{tab_results_bayes_dust}
  \begin{tabular}{lcccc}
  \hline
 \bfseries {ID} &$\mathbf{\alpha_{dust}}$ & $\mathbf{\beta_{dust}}$  & $\mathbf{T_{dust}}$ & $\mathbf{L_{dust}}$\\
    &  &  &  [K] & [$10^{12}$ L$_\odot$]\\
 \hline
  UDF1  & $1.80\pm0.09$ &  $2.0\pm0.3$ &  $56\pm3$ & $3.5\pm0.2$ \\
  UDF3  & $2.2\pm0.1$ &  $2.1\pm0.3$ &  $73\pm4$ & $4.9\pm0.3$ \\
  UDF5  & $1.80\pm0.09$ & $2.3\pm0.3$ &  $42\pm3$ & $0.77\pm0.04$ \\
  UDF8  & $1.60\pm0.08$ & $2.2\pm0.3$ & $52\pm4$ & $1.10\pm0.06$ \\
  UDF10 & $1.9\pm0.1$ & $2.1\pm0.3$ & $46\pm7$ & $0.41\pm0.05$ \\
  UDF11 & $2.2\pm0.1$ & $2.5\pm0.2$ & $69\pm4$ & $2.2\pm0.2$ \\
  UDF13 & $1.8\pm0.1$ & $2.3\pm0.3$ & $60\pm3$ & $1.2\pm0.2$\\
  ALESS067.1 & $1.80\pm0.09$ & $2.2\pm0.2$ & $50\pm3$ & $5.4\pm0.3$\\
  AzTEC.GS25 & $1.80\pm0.09$ &  $2.3\pm0.2$ & $40\pm2$ & $3.9\pm0.2$\\
  AzTEC.GS21 & $2.2\pm0.1$ &  $1.8\pm0.2$ & $63\pm3$ & $3.9\pm0.2$\\ 
  AzTEC.GS22 & $1.9\pm0.1$ &  $1.8\pm0.3$ & $40\pm2$ & $1.01\pm0.06$\\ 
  \hline
  \end{tabular}
\end{table}

\section{Results}\label{sec_add_info} 

In this Section we present and comment the information extracted from galaxy multi-wavelength emission. In Fig.~\ref{panel} (left column) we show the best SEDs (thick solid black lines) for the 11 (sub-)millimetre selected star-forming galaxies of our sample, as obtained by the broad-band SED fitting performed with CIGALE. The contribution of each component to the total SED is color-coded and comprises: the UV/optical/NIR emission coming from young stars (i.e., enshrouded in their BC, see Sect.~\ref{sec_stellar_component}); the UV/optical/NIR emission of old stars (i.e., in the galaxy ISM, outside their BC, see Sect.~\ref{sec_stellar_component}); the warm dust MID power-law and the cold dust FIR modified BB \citep[][see Sect.~\ref{sec_dust_emission}]{Casey2012:2012MNRAS.425.3094C}. The spectroscopic redshift of each source and the reduced $\chi^2$ (i.e., $\chi_{\rm red}^2=\chi^2/{\rm dof}$) are written on the corresponding panel. 
For reference we add the predicted radio emission coming solely from the host galaxy star formation (solid yellow line; its derivation will be described in Sect.~\ref{sec_radio_emission}). The observed fluxes in the radio band (at $\nu_{\rm obs}=6$ GHz, 1.4 GHz, 610 MHz; yellow filled circles) are consistent with the yellow solid line (within the uncertainties). We do not include radio photometry in SED fitting since we want to avoid any strong assumption on the radio-to-(sub)mm spectral index. The latter is well constrained when at least two fluxes in the radio band are available, but this requirement is satisfied only for one source, i.e. ALESS067.1.

The main physical quantities characterizing our galaxies (e.g., SFR, $\tau_\star$, M$_{\rm star}$, T$_{\rm dust}$) are obtained both with $\chi^2$ analysis and bayesian analysis. The latter approach estimates galaxy physical properties from likelihood-weighted parameters on a fixed grid of models (whose set up for this work has been described in the previous Sections), exploring the parameter space around the given priors (listed in Tab. \ref{tab_cigale_param_priors}). We expect this analysis to provide the most precise and accurate estimations of the main SED-inferred quantities and we list the corresponding outcomes in Tabs. \ref{tab_results_bayes_stars} and \ref{tab_results_bayes_dust}, along with their uncertainties. We note that these quantities are well constrained only if the probability distribution function (pdf) is well behaved (e.g., single peak). This is the case for the majority of the marginalized pdfs. The most common exceptions are found for: the BC-to-ISM V-band attenuation, $f_{\rm att}$; the BC attenuation spectral index, $\delta^{\rm BC}$; the separation age between the old and young SSPs, $\Delta_\star$.
Finally, the $\chi^2$ analysis have been exploited to have a hint on fits goodness, i.e. $\chi^2_{\rm red} \sim 1$ (cfr. Tab. \ref{tab_results_bayes_stars}, second column). 

Given the outcomes listed in Tabs. \ref{tab_results_bayes_stars} and \ref{tab_results_bayes_dust}, in the following we discuss the resulting attenuation laws for our objects (Sect.~\ref{sec_att_law}), we derive their dust and gas masses (Sects.~\ref{sec_dust_mass} and \ref{sec_gas_mass}) and provide the analysis on their X-ray and radio emission (Sects.~\ref{sec_Xray_analysis} and \ref{sec_radio_emission}). Finally, we collect from literature the multi-wavelength sizes of our objects and convert them to the corresponding circularized radii (Sect.~\ref{sec_multiwavelength_sizes}).

\begin{figure*}
  \centering
\includegraphics[width=2\columnwidth]{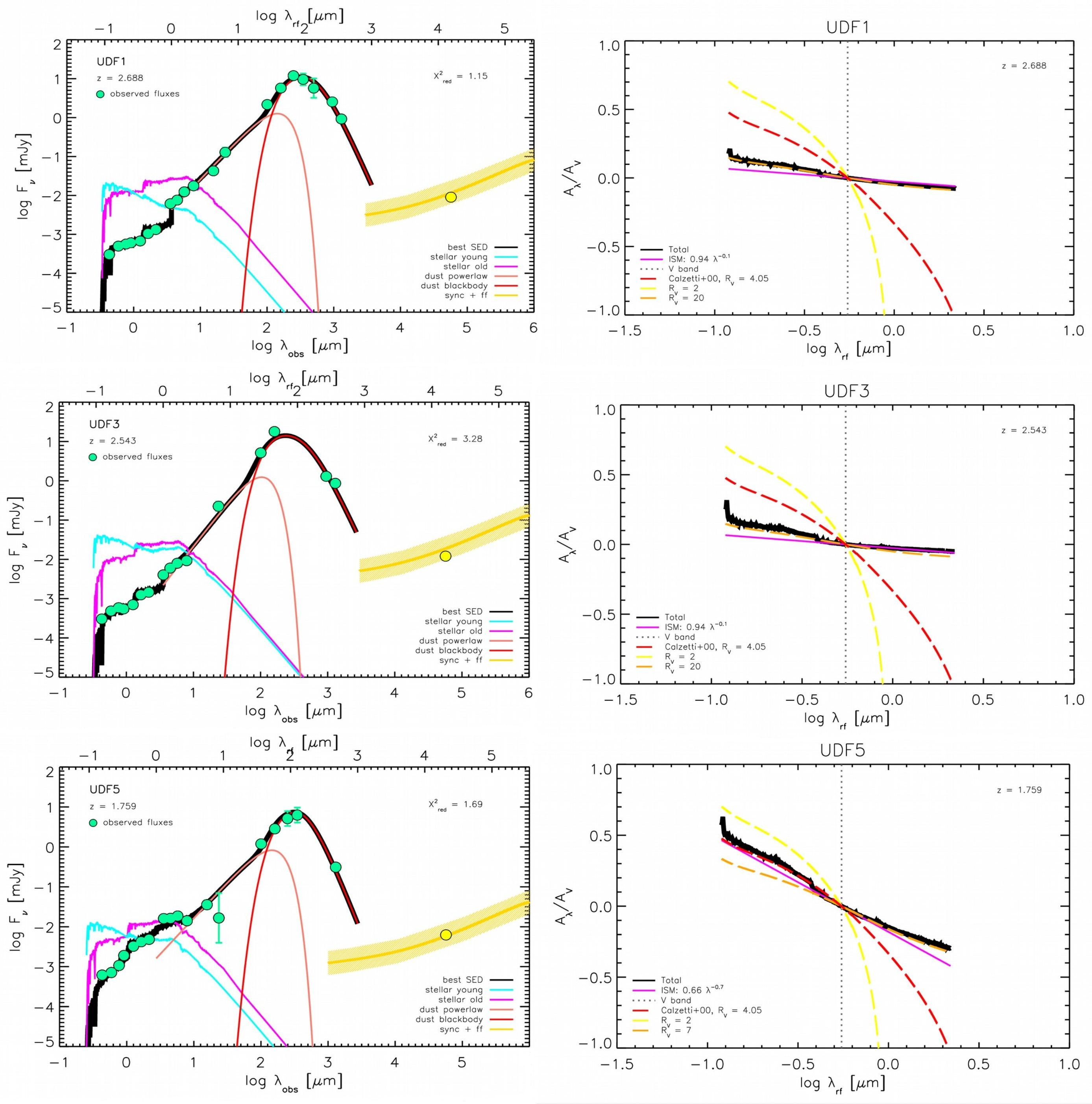} 
\caption{Galaxy best SEDs (left column) and best attenuation laws (right column). Left panels: thick solid black line stands for the total best galaxy SED; solid cyan line indicates the UV/optical/NIR emission coming from young stars; solid magenta line stands for the old stars UV/optical/NIR emission; solid orange line represents the warm dust MID power-law; solid red line shows the cold dust FIR modified BB emission; thick solid yellow line represents the radio emission coming solely from host galaxy star formation; green filled circles are the observed fluxes included in SED fitting and yellow filled circles are the observed fluxes in the radio bands (not included in the fit). Error bars are omitted for clarity when they are comparable to the symbol size. Right panels: thick solid black line stands for the total best attenuation law; solid magenta line is the ISM attenuation law; dashed red line represents the standard Calzetti attenuation law; dashed yellow and orange lines are the Calzetti attenuation laws obtained by decreasing (yellow) or increasing (orange) the standard value for R$_{\rm V}$, as it is indicated in the legends.}\label{panel}
\end{figure*}
\begin{figure*}
  \centering
\includegraphics[width=2\columnwidth]{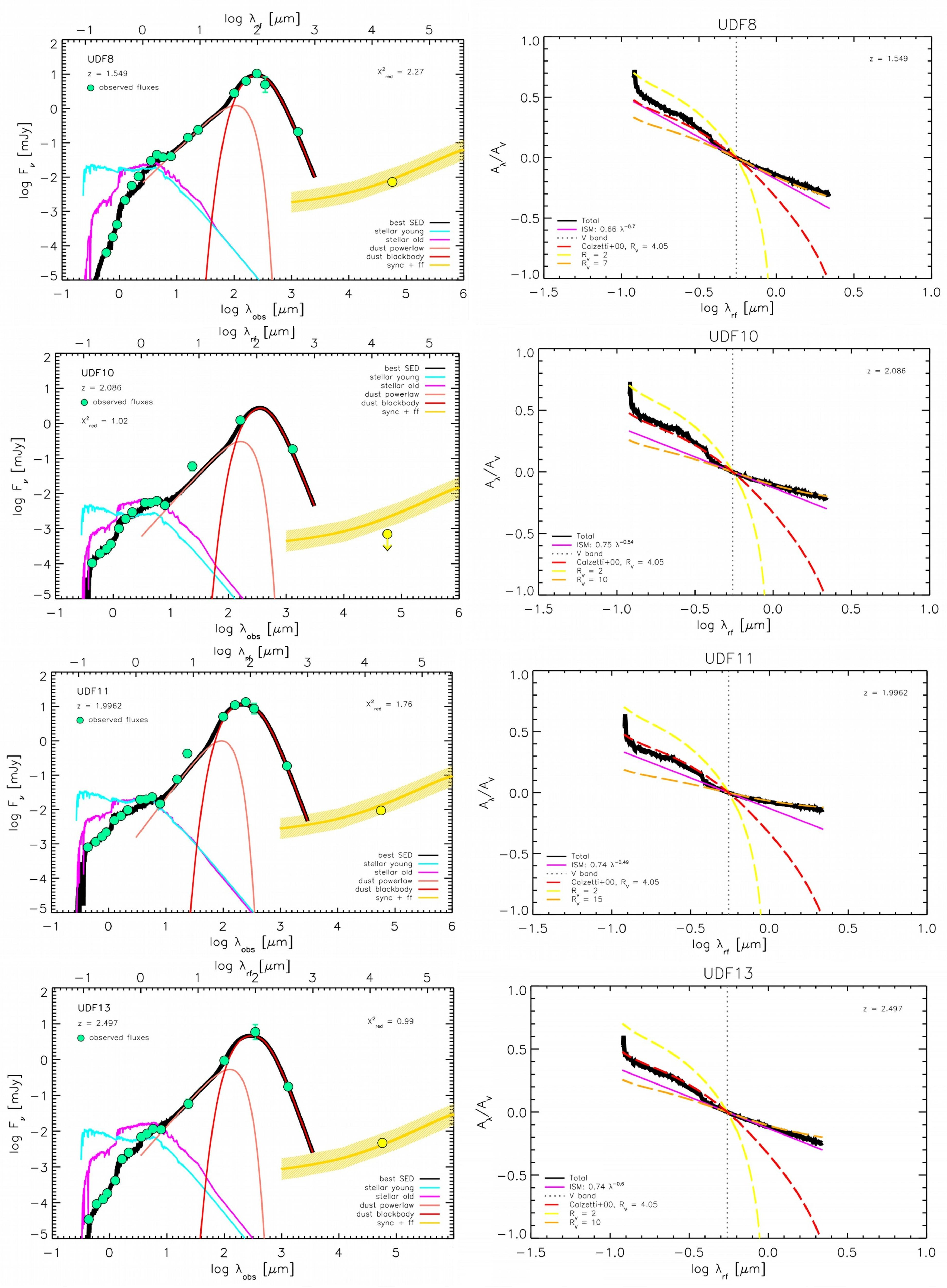}
\contcaption{}
\end{figure*}
\begin{figure*}
  \centering
\includegraphics[width=2\columnwidth]{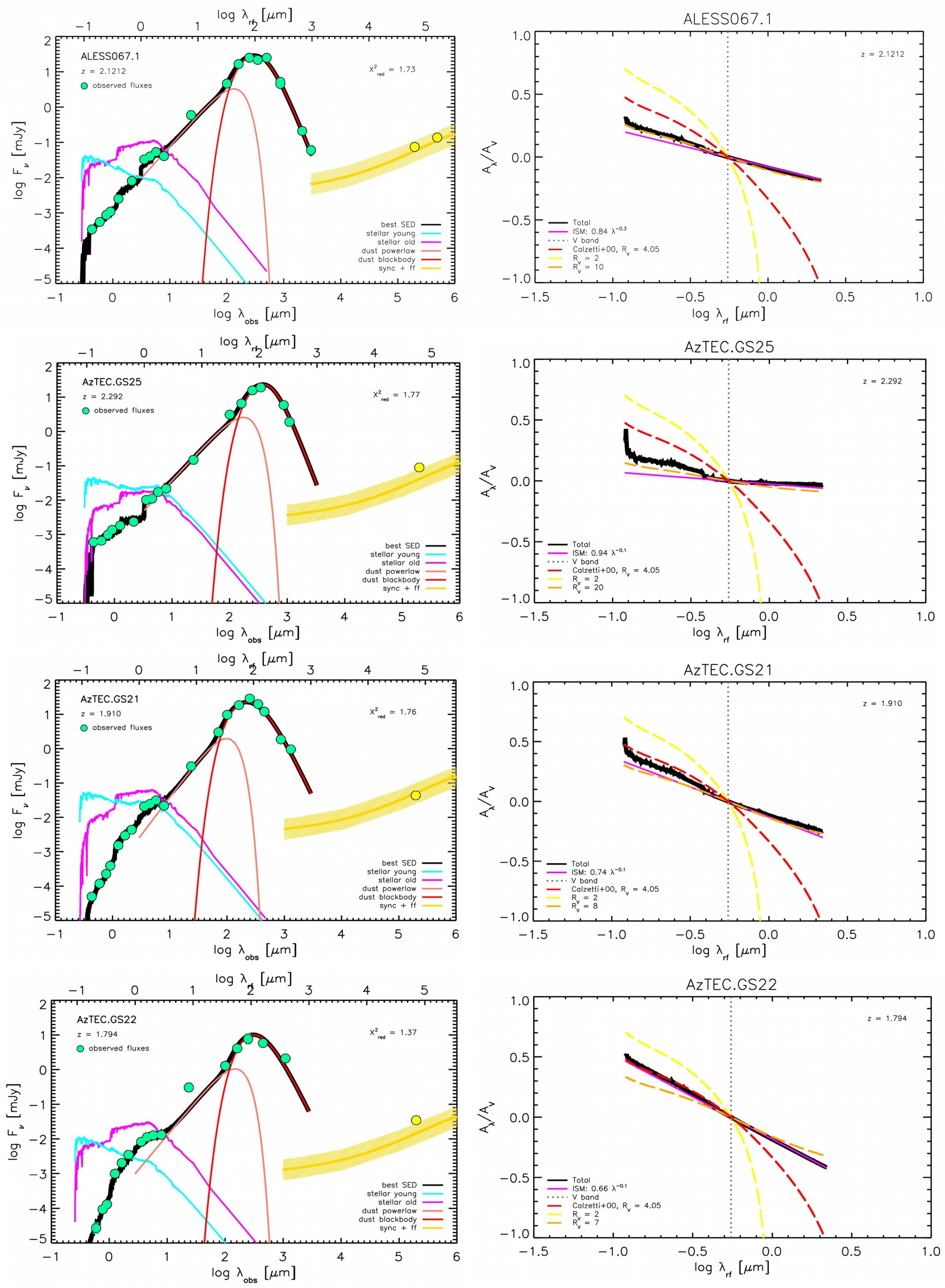}
\contcaption{}
\end{figure*}

\subsection{Attenuation law}\label{sec_att_law}

In Fig.~\ref{panel} we show the attenuation law (i.e., $A_{\rm \lambda}/A_{\rm V}$; thick solid black line) obtained as described in Appendix \ref{appendix_attlaw} by adopting the prescriptions already explained in Sect.~\ref{sec_dust_att}. 

We note that the overall emerging stellar emission is mostly shaped by dust attenuation in the ISM ($\sim \lambda^{-\delta^{\rm ISM}}$, with $0.1\leq \delta^{\rm ISM}\leq 0.7$; solid magenta line). Indeed, we modelled dust extinction in BCs to absorb almost all the radiation from young stars enshrouded in these dense environments. Some emission from young stars emerges just at $\lambda_{\rm rest}\gtrsim 10$ $\mu$m, where dust extinction is less effective.

Focusing on the total (i.e., ISM + BC) attenuation law ($A_{\lambda}/A_{\rm V}$), we note two main different behaviours in $\lambda$. For the majority of our galaxies (seven out of eleven sources) the attenuation law is well described by a standard \citet[i.e. $R_{\rm V}=A_{\rm V}/{\rm E(B-V)}\simeq 4.05\pm0.80$; dashed red line]{Calzetti2000:2000ApJ...533..682C}  at wavelengths bluer than $\lambda_{\rm V}=540$ nm, while it shows a flattening towards redder wavelengths, with a characteristic $R_{\rm V}$ ranging between 7 and 15 (crf. with dashed orange and yellow line). The other five objects show a flatter attenuation law at every $\lambda$, with a $R_{\rm V}\simeq20$.  
This flattening of the attenuation law towards high-z (Calzetti law is constrained in local star-forming galaxies like Milky Way) may be caused by a more uniformly mixed geometry of the interstellar dust grains or it may simply indicate diverse dust grains geometries and distributions into the ISM, that are very difficult to be constrained at $z>1$ \citep[e.g.,][]{Salmon2016:2016ApJ...827...20S, Leja2017:2017ApJ...837..170L, Narayanan2018:2018ApJ...869...70N, Salim2018:2018ApJ...859...11S, Buat2019:2019A&A...632A..79B, Trayford2019:2019MNRAS.485.5715T}. 
Because of the smaller amount of reddening at near-infrared wavelengths, the standard Calzetti has been found to significantly lower SFR and stellar mass when applied to high-z dusty galaxies \citep[e.g.,][]{Williams2019:2019ApJ...884..154W}. Exploiting such a locally-calibrated attenuation law at high-z should be done with caution, in particular while dealing with the UV/optical-to-millimetre emission of DSFGs.

\subsection{Dust mass}\label{sec_dust_mass}
 \begin{table*}
 \centering
  \caption{For each source of the sample (ID in column 1) we list of the values for the quantities entering in Eq. \ref{eq_dust_mass} (values for $\beta_{\rm dust}$ and T$_{\rm dust}$ can be found in Tab. \ref{tab_results_bayes_dust}) and the resulting dust mass (M$_{\rm dust}$; last column, in units of $10^8$ M$_\odot$). In particular, $\nu_{\rm dust}^{\rm rest}$ ($\equiv\lambda_{\rm dust}^{\rm rest}$>200 $\mu$m) is the rest-frame frequency corresponding to the observed R-J flux S$_{\rm dust}\equiv$ S($\nu_{\rm obs}$): $a$) ALMA flux at $\lambda_{\rm obs}=1300$ $\mu$m ($\nu_{\rm obs}=230$ GHz); $b$) ALMA flux at $\lambda_{\rm obs}=870$ $\mu$m ($\nu_{\rm obs}=345$ GHz); $c$) AzTEC flux at $\lambda_{\rm obs}=1100$ $\mu$m ($\nu_{\rm obs}=273$ GHz). Units are reported between square brackets. Dust masses are not corrected by \citet[]{Magdis2012;2012ApJ...760....6M}: a factor 2 must be added.}\label{tab_dust_mass}
  \begin{tabular}{lccccc}
\hline
 \bfseries {ID}  & $\mathbf{d_L}$ &$\mathbf{\lambda_{dust}^{rest}}$&
$\mathbf{\nu_{dust}^{rest}}$&$\mathbf{S_{ dust}}$ & $\mathbf{M_{ dust}}$  \\
 & [Mpc] & [$\mu$m]&[GHz]& [mJy] & [$10^{8}\,M_\odot$]\\
 \hline
  UDF1  &22886.4& 353 & 850& $0.924\pm0.076^a$ & $2.8\pm1.0$ \\
  UDF3    & 21294.4 & 367 & 817& $0.863\pm0.084^a$& $2.0\pm0.8$ \\ 
  UDF5    & 13553.1 & 471 &637 & $0.311\pm0.049^a$ & $2.3\pm1.1$ \\
  UDF8    &11588.6& 510& 588 &$0.208\pm0.046^a$ & $1.21\pm0.67$ \\
  UDF10  &16711.9 & 421&  712 & $0.184\pm0.046^a$& $0.95\pm0.67$ \\
  UDF11  &15833.4&  434& 691& $0.186\pm0.046^a$ & $0.73\pm0.33$ \\
  UDF13  &20825.4& 372& 806& $0.174\pm0.045^a$ & $0.60\pm0.34$ \\
  ALESS067.1  &17058.3& 279& 1075& $4.5\pm0.4^b$ & $4.8\pm1.8$ \\
  AzTEC.GS25  &18755.5& 334& 898& $1.9\pm0.6^c$ & $6.8\pm4.1$ \\
  AzTEC.GS21  & 14997.8& 445 & 674& $0.954\pm0.074^a$ & $2.9\pm0.7$ \\
  AzTEC.GS22   &13885.7& 394 & 761& $2.1\pm0.6^c$ & $7\pm4$ \\
  \hline
  \end{tabular}
  \end{table*}

We derive the dust mass for each source, given a measure of dust flux in the rest-frame Rayleigh-Jeans (R-J) regime and the bayesian estimation of dust temperature. It is well known \citep[e.g.,][]{Bianchi2013:2013A&A...552A..89B, Gilli2014:2014A&A...562A..67G, D'Amato2020:2020A&A...636A..37D, Pozzi2020:2020MNRAS.491.5073P} that dust mass can be estimated in the optically thin approximation ($\tau_\nu \ll 1$) as:
\begin{equation}\label{eq_dust_mass}
 M_{\rm dust} = \frac{S(\nu_{\rm obs})\, d_{\rm L}^2}{(1+z)\, k(\nu_{\rm rest})\, B_{\rm BB}(\nu_{\rm rest},T_{\rm dust}) }
\end{equation}
where $S(\nu_{\rm obs})$ is the observed flux such as $\nu_{\rm rest}=\nu_{\rm obs}(1+z)$ is in R-J regime; $B_{\rm BB}$ is the BB brightness computed at $\nu_{\rm rest}$, with $T_{\rm dust}$ being the dust equilibrium temperature derived by performing the single-T fit of the FIR SED; $k(\nu_{\rm rest}) = 5.1\,(\nu_{\rm rest}/1.2\,{\rm THz})^\beta$ cm$^2$ g$^{-1}$ is the dust absorption coefficient per unit of mass \citep[e.g.,][]{Magdis2012;2012ApJ...760....6M, Gilli2014:2014A&A...562A..67G}; $d_{\rm L}^2$ is the luminosity distance; $(1+z)$ takes into account the k-correction entering in the relation between flux and luminosity. 

In order to obtain the most reliable estimate of dust mass we exploited the reddest (sub-)millimetre observed flux for each source. In Tab. \ref{tab_dust_mass} we list the corresponding rest-frame wavelength, that lies to a good extent in the R-J regime. We estimated errors on dust mass following error propagation theory. However, dust masses derived from single-temperature fit to the FIR SED are \textit{luminosity-weighted}. As a consequence, we select a dust temperature that is typically slightly higher than the mean and the resulting dust mass tends to be underestimated. \citet{Magdis2012;2012ApJ...760....6M} attempted to quantify this effect, which turns out to shift downwards dust masses of a factor $\sim2$ (cfr. Sects. \ref{sec_dust_emission} and \ref{sec_gas_mass}). 

\subsection{Gas mass}\label{sec_gas_mass}

 \begin{table}
 \centering 
 \caption{CO analysis: $L_{\rm CO}$ and $M_{\rm H_2}$ for UDF1, UDF3 UDF8 and ALESS067.1 by Pantoni et al. (in preparation), assuming an $\alpha_{\rm CO}=3.6$ K km pc$^2$ s$^{-1}$ M$_{\rm \odot}^{-1}$.}\label{tab_decarli}
  \begin{tabular}{lcccc}
  \hline
 $\mathbf{ID}$   & $\mathbf{z_{CO}}$  & \bfseries {CO line} & $\mathbf{L_{CO}}$  & $\mathbf{M_{H_2}}$ \\
 &  &  & [${10}^8$ K km s$^{-1}$ pc$^2$] &[$10^{10}$ M$_\odot$] \\
 \hline
 UDF1 & 2.698 & J(3-2) &  $31\pm3$ & $2.6\pm0.7$ \\
 UDF3 & 2.543 & J(3-2) &  $170\pm10$ & $15\pm3$  \\
 UDF8 & 1.5490 & J(2-1) & $122\pm9$ & $5.8\pm1.1$ \\
 ALESS067.1 & 2.1212 & J(3-2) & $196\pm31$ & $16.8\pm5.4$ \\
  \hline
  \end{tabular}
\end{table}
CIGALE does not allow to derive galaxy gas mass directly from broad band fitting: gas masses computed by the stellar emission module consist solely of the gas fraction that is restituted to the medium by stellar evolution (i.e., M$_{\rm R}$; see Tab. \ref{tab_results_bayes_stars}). We derive the gas masses for our 11 DSFGs by relying either on CO line luminosities (when available) or on R-J interstellar dust continuum.

Although CO lines require to assume a conversion factor $\alpha_{\rm CO}$ (to convert the observed CO line luminosity into the galaxy molecular hydrogen mass, $H_2$) and an excitation ladder (in case of transitions with $J>1$), they provide the most direct method to infer the molecular gas content in high-z DSFGs.

In Tab. \ref{tab_decarli} we list the CO-derived $H_2$ masses for four sources of our sample (UDF1, UDF3, UDF8, ALESS067.1)  by Pantoni et al. (in preparation), who use the same approach, tool and prescriptions adopted in this work. We note that the $H_2$ mass, when multiplied by a factor 1.36 that accounts for Helium, provides the total molecular gas mass of the galaxy with a very good approximation. We derive the molecular hydrogen mass from $J>1$ CO line luminosities, converted to the equivalent CO(1-0) luminosity following \citet{Daddi2010:2010ApJ...713..686D, Daddi2015:2015A&A...577A..46D}, by assuming the same excitation ladder, i.e. $r_{31}=0.42\pm0.07$ and $r_{21}=0.76\pm0.09$, and CO conversion factor, i.e. $\alpha_{\rm CO} = 3.6$ M$_\odot$(K km s$^{-1}$ pc$^2$ )$^{-1}$. Although the topic is still under debate, a smaller value of $\alpha_{\rm CO}\sim0.8-1$ (often exploited for local ULIRGs and high-z compact DSFGs) is commonly thought to be inappropriate for the globally distributed molecular gas and more advisable for high-resolution observations which isolate the nuclear region \citep[e.g., ][for a review]{Scoville2016:2016ApJ...820...83S, Carilli2013:2013ARA&A..51..105C}.

Observing CO lines at high-z is very expensive in terms of time-on-source ($\gtrsim$ a few hours) and so they are available just for a small number of 
DSFGs. Alternative methods are mainly based on the exploitation of continuum far-IR thermal emission coming from cold interstellar dust but they are affected by larger uncertainties since they need to assume a gas-to-dust ratio (GDR). Nevertheless, they are very convenient for high-z massive dusty galaxies: indeed their dust continuum can be detected, e.g. by ALMA, in just a few minutes of observing time. The two most popular methods exploiting dust R-J continuum are developed and described in the articles by \citet{Leroy2011:2011ApJ...737...12L} and \citet{Scoville2014:2014ApJ...783...84S, Scoville2016:2016ApJ...820...83S}, to which we refer for the following analysis. We note that these two approaches have been recently combined in the work by \citet{Liu2019:2019ApJS..244...40L} on the A3 COSMOS\footnote{https://sites.google.com/view/a3cosmos} sample.

We derived the total gas mass, i.e. $HI+H_2$, using the local relation by \citet[their Sect. 5.2]{Leroy2011:2011ApJ...737...12L}:
\begin{equation}\label{eq_leroy}
\begin{split}
    \log_{10}&({\rm GDR}) = \log_{10} \frac{M_{\rm HI}+M_{{\rm H}_2}}{M_{\rm dust}}\\ 
   & = (9.4\pm1.1)-(0.85\pm0.13)[12+\log_{10}(O/H)]. 
\end{split}
\end{equation}
The dependence on gas metallicity allows to extend the result to $z\lesssim3$. We derived the gas metallicity for our $z\sim2$ sample of DSFGs using the mass-metallicity relation by \citet[their Sect. 2.2]{Genzel2012:2012ApJ...746...69G}, following \citet[their Sect. 2.4]{Elbaz2018:2018A&A...616A.110E}:
\begin{equation}\label{eq_elbaz}
 12+\log_{10}(O/H)= -4.51+2.18\, M_{\star}-0.0896\,(\log_{10}M_{\star})^2.
\end{equation}
The rms dispersion of mass-metallicity relation at $z\sim2$ is of about $0.09$ dex. Systematic uncertainties between different metallicity indicators and calibrators can reach $\sim0.3$ dex \citep[e.g.,][]{Kewley2008:2008ApJ...681.1183K} and clearly dominate over the statistical one. The outcomes are shown in Fig.~\ref{fig_gas_metallicity_elbaz}.
In order to derive the total gas mass (Eq. \ref{eq_leroy}), we used our SED-inferred dust masses (crf. Tab. \ref{tab_dust_mass}) corrected following \citet{Magdis2012;2012ApJ...760....6M}. As discussed in Sect.~\ref{sec_dust_mass}, this procedure brings into the total dust mass budget also the coldest interstellar dust, whose content is (on average) a factor $\sim 2$ underestimated when a single-temperature modified BB is used to fit the FIR interstellar dust thermal emission.
The resulting total gas masses M$_{\rm gas,\,tot}$ are listed in Tab. \ref{tab_gas_mass_results}. The uncertainty is of about 0.3 dex. We note that metallicities could be even higher, in case of very compact FIR sources ($r_{\rm FIR}<1$ kpc) and possibly dust thick. Actually the inferred total gas masses strongly depend on the method and, in our case, on the assumed mass-metallicity relation, shown in Eq.~\eqref{eq_elbaz}.

We estimated the molecular ISM masses of our 11 DSFGs using the empirical calibration by \citet[cfr. their Fig. 1, right panel]{Scoville2016:2016ApJ...820...83S}:
\begin{equation}
 \frac{L_{\nu_{850\,\mu m}}}{M_{\rm gas,\, mol}} = 6.2\times10^{19}\biggl(\frac{L_{\nu_{850\,\mu m}}}{10^{31}}\biggr)^{0.07}.
\end{equation}
The resulting molecular gas masses M$_{\rm gas,\,mol}$ are listed in Tab. \ref{tab_gas_mass_results}. The corresponding uncertainty is of about 0.3 dex.
We note that our two estimates for gas mass (i.e., M$_{\rm gas,\,mol}$ and M$_{\rm gas,\,tot}$) are compatible within the errors.

 \begin{table} 
\centering 
\caption{Molecular gas mass M$_{\rm gas,\,mol}$ derived following the approach in \citet{Scoville2016:2016ApJ...820...83S}; total (HI+H$_2$) gas mass M$_{\rm gas,\,tot}$ and gas metallicity Z$_{\rm gas}$ evaluated following \citet{Genzel2012:2012ApJ...746...69G} and \citet{Elbaz2018:2018A&A...616A.110E}. Uncertainties have been omitted for clarity: they are $\sim0.3$ dex for molecular gas masses and $\sim0.4$ dex for total gas masses and gas metallicities.}\label{tab_gas_mass_results}
\begin{tabular}{lcccc}
\hline
 \bfseries {ID}  & $\mathbf{log\,M_{gas,\,mol}}$ & $\mathbf{log\,M_{gas,\,tot}}$  & $\mathbf{GDR}$  & $\mathbf{Z_{gas}}$ \\ 
  & [M$_\odot$] & [M$_\odot$]& &  12+log(O/H) \\ 
  \hline
  UDF1 & 10.5 & 10.8 & 120 & 8.61 \\ 
  UDF3 & 10.6 & 10.7 & 119 & 8.62 \\ 
  UDF5 & 10.1 & 10.9 & 161 & 8.46 \\ 
  UDF8 & 10.0 & 10.5 &  126 & 8.59 \\
  UDF10 & 9.8 & 10.5 & 159 & 8.47 \\ 
  UDF11 & 9.8 & 10.3 & 127 & 8.58 \\ 
  UDF13 & 9.7 & 10.2 & 127 & 8.59 \\ 
  ALESS067.1 & 10.9 & 11.0 & 100 & 8.71 \\
  AzTEC.GS25 & 10.6 & 11.2 & 121 & 8.61 \\ 
  AzTEC.GS21 & 10.7 & 10.8 & 106 & 8.68 \\ 
  AzTEC.GS22 & 10.7 & 11.3 & 130 & 8.57 \\ 
  \hline
  \end{tabular}
 \end{table} 

\subsection{X-ray emission}\label{sec_Xray_analysis}

The X-ray emission from a star-forming galaxy can be traced back to two main different processes: the star-formation itself, since massive, compact binaries can produce X-ray radiation (the so called \textit{X-ray binaries}), and the accretion of matter onto the central SMBH (if the AGN is \textit{on}), since the infalling heated material radiate (also) in the X-ray band. Thus, the X-ray luminosity of a star-forming galaxy can provide a wealth of information on the possible presence of a central AGN and on its evolutionary stage.

As described in Sect.~\ref{sec_sample}, nine sources of our sample out of eleven own a robust counterpart in the recent and very deep X-ray \textit{Chandra} catalog by \citet{Luo2017:2017ApJS..228....2L}, based on a $\approx$ 7 Ms map of the CDF-S. We note that the two X-ray non-detections (UDF5 and AzTEC.GS22) lie in very deep regions of the \textit{Chandra} map (equivalent exposure times are $\sim6.22$ Ms and $\sim5.80$ Ms, respectively), thus the hypothesis of totally obscured X-ray sources is the most probable.

For every X-ray source, \citet{Luo2017:2017ApJS..228....2L} provide the $0.5-7.0$ keV \textit{intrinsic} luminosity, i.e. net of the Milky Way and X-ray source intrinsic absorption, the latter determined by the intrinsic column density $N_{\rm H,int}$. In Appendix \ref{appendix_luo} we calculate the corresponding $2-10$ keV luminosity, in order to easily compare with literature while investigating galaxy-BH co-evolution. Intrinsic $2-10$ keV luminosities ($L_{\rm X}$) for our nine sources with an X-ray counterpart are listed in Tab. \ref{tab_luo}.

\begin{table*}
\centering
 \caption{In this table we list: IDs of the source associations (ID: this work; ID$_{\rm X}$: Luo et al. 2017) and their angular separation (d) in arcsec; source redshifts (z); $2-10$ keV intrinsic luminosities at redshift $z$ ($L_{\rm X}$); the class (AGN or galaxy) associated to each source by \citet{Luo2017:2017ApJS..228....2L} and the X-ray dominant component (active nucleus or host galaxy) found by our analysis.}\label{tab_luo}
  \begin{tabular}{lccccccc}
 \hline
 \bfseries {ID} & $\mathbf{ID_X}$ & \bfseries {d} &$\mathbf{z}$ & $\mathbf{L_{X}}$  & \bfseries{class X}& \bfseries{X dominant} \\
  & & [arcsec] & & [$10^{42}$ erg s$^{-1}$] & Luo et al.&\bfseries{component}\\
 \hline
  UDF1 & 805 & 0.69 & 2.698 & 40.2 & AGN & AGN \\
  UDF3 & 718 & 0.54 & 2.544 & 1.8 & AGN & galaxy \\
  UDF8 & 748 & 0.07 & 1.549 & 36.3 & AGN & AGN \\
  UDF10 & 756 & 0.31 & 2.086 & 0.6 & galaxy & galaxy\\
  UDF11 & 751 & 0.29 & 1.996 & 1.7 & galaxy & galaxy \\
  UDF13 & 655 & 0.26 & 2.497 & 2.1 & AGN & galaxy\\
  ALESS067.1  & 794 & 0.40 & 2.1212 & 3.8 & AGN & galaxy\\
  AzTEC.GS25  & 844 & 0.71 & 2.292 & 6.1 & AGN & galaxy\\
  AzTEC.GS21 & 852 & 0.36 & 1.91 & 1.7 & AGN & galaxy\\
  \hline
  \end{tabular}
\end{table*}

\subsubsection{X-ray dominant component}
\citet{Luo2017:2017ApJS..228....2L} provide also a classification of the X-ray sources (i.e., AGN, galaxy or star), that we show for reference in Tab. \ref{tab_luo} (column 12 - \textbf{classX} - Luo et al.). 
\begin{table*}
\centering
 \caption{In this table we list: IDs of the sources; $2-10$ keV luminosities ($L_{\rm X}$); total infrared luminosities ($8-1000$ $\mu$m rest frame) obtained from SED fitting ($L^{\rm dust}_{\rm IR}$, as described in Sect.~\ref{sec_dust_emission}); infrared AGN luminosities and AGN fractions ($f_{\rm AGN}^{(1)}$) inferred following Mullaney et al. (2011; equation \ref{Eq_Mullaney_fAGN}); AGN fraction inferred by using \citet{Asmus2015:2015MNRAS.454..766A} MIR-X ray correlation ($f_{\rm AGN}^{(2)}$).}\label{tab_mullaney}
  \begin{tabular}{lccccc}
 \hline
 \bfseries {ID} & $\mathbf{L_{\rm X}}$  & $\mathbf{L^{\rm dust}_{\rm IR}}$ & $\mathbf{L^{\rm AGN}_{\rm IR}}$ & $\mathbf{f_{\rm AGN}^{(1)}}$ & $\mathbf{f_{\rm AGN}^{(2)}}$\\
  & [$10^{43}$ erg s$^{-1}$]  & [$10^{43}$ erg s$^{-1}$] & [$10^{43}$ erg s$^{-1}$] & [\%] & [\%]\\
 \hline
 UDF1 & $4.0\pm0.4$ & $1345\pm67$ & $16\pm13$ & 1 & 6\\
 UDF3 & $0.18\pm0.02$ & $1910\pm114$ & $0.5\pm0.3$ &  0.03 & 0.2\\
 UDF8 & $3.6\pm0.4$ & $430\pm21$ & $14\pm11$ & 3 & 14 \\
 UDF10 & $0.060\pm0.006$ & $159\pm19$ & $0.15\pm0.07$ & 0.09& 1\\
 UDF11 & $0.17\pm0.02$ & $873\pm59$ & $0.5\pm0.3$ & 0.06 & 0.5\\
 UDF13 & $0.21\pm0.02$ & $458\pm75$ & $0.6\pm0.4$  & 0.1 & 0.8\\
 ALESS067.1 & $0.38\pm0.04$ & $2096\pm105$ & $1.2\pm0.7$ & 0.06 & 0.4\\
 AzTEC.GS25 & $0.61\pm0.06$ & $1493\pm75$ & $2\pm1$ & 0.1 &1\\
 AzTEC.GS21 & $0.17\pm0.2$ & $1546\pm77$ & $0.5\pm0.3$ & 0.03 & 0.3\\
   \hline
  \end{tabular}
\end{table*}
They classified as \textit{X-ray AGN} every X-ray source that shows at least one indication on the presence of a central AGN emitting in the  X-rays (for a detailed description of the criteria they used we refer the reader to their Sect. 4.5 and references therein). However, it does not imply that the AGN X-ray emission prevails over the galactic one. 

\begin{figure}
  \centering
\includegraphics[width=\columnwidth]{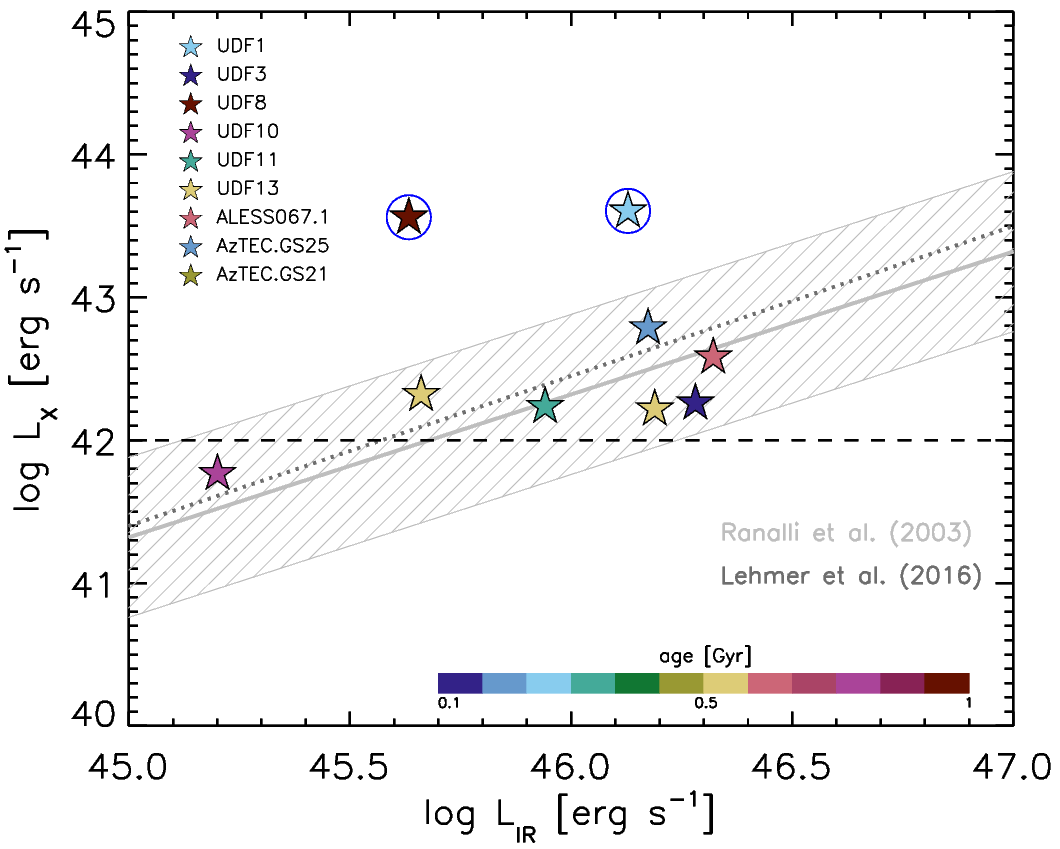}
\caption{X-ray luminosity versus IR luminosity. Stars stand for the sources with an X-ray counterpart in the catalog by \citet{Luo2017:2017ApJS..228....2L}. They are color-coded by age (i.e., $\tau_\star$). IR luminosities ($8-1000$ $\mu$m rest-frame) result from SED fitting. Gray line represents the correlation between X-ray and IR luminosity when they are ascribed to star formation solely by \citet{Ranalli2003:2003A&A...399...39R}, with its 1$\sigma$ scatter. Dark gray dotted line represents the trend followed by our objects when the evolution with galaxy $z$, M$_\star$ and SFR is taken into account in the relation by \citet{Lehmer2016:2016ApJ...825....7L}. Its 1$\sigma$ scatter (of about 0.4 dex) has not been plotted for clarity. The two outliers are highlighted with a blue circle. The dashed horizontal black line indicates the value at which X-ray luminosity from the central BH begins to be comparable to the one coming from star formation.}\label{fig_ranalli_age}
\end{figure}

To get a deeper insight on this topic, in Fig.~\ref{fig_ranalli_age} we compare the $2-10$ keV luminosity of our sources with their infrared luminosity (derived from SED fitting, see Tab. \ref{tab_results_bayes_dust}). We note every source falling above the relation by \citet{Ranalli2003:2003A&A...399...39R} for star-forming galaxies to be an AGN, meaning that its X-ray luminosity overwhelms the one coming from the host galaxy. In Fig.~\ref{fig_ranalli_age} these sources are marked with a blue circle. This result is still valid when the evolution of X-ray binaries luminosity with galaxy redshift, SFR and M$_\star$ is taken into consideration \citep[e.g.,][]{Lehmer2016:2016ApJ...825....7L}. Note that the two \textit{AGN-dominated} X-ray sources are also in the 3Ms XMM catalog by \citet{Ranalli2013:2013A&A...555A..42R}.

Actually, many recent works have shown that the X-ray luminosity from the central AGN (if present) begins to be comparable to the host galaxy one at $L_{\rm X}\approx 10^{42}$ erg s$^{-1}$ \citep[e.g.,][]{Bonzini2013:2013MNRAS.436.3759B, Padovani2015:2015MNRAS.452.1263P}. Indeed, this value is commonly adopted to clearly discern the nuclear X-ray emission from that associated to star formation $L_{\rm X,\,SFR}\approx7\times10^{41}$ erg s$^{-1}$ SFR/$10^2$ M$_\odot$ yr$^{-1}$ \citep[e.g.,][]{Vattakunnel2012:2012MNRAS.420.2190V}. In Fig.~\ref{fig_ranalli_age} we represent this threshold with a dashed black line. We note that almost all the sources with an X-ray counterpart lie above it, possibly indicating that a X-ray quasar is growing in the nuclear region of the host galaxy. 
We refer the reader to Sect. \ref{sec_results_discussion} for a further analysis on this topic.

\subsubsection{AGN fraction in the IR domain}
In the following we exploit the $2-10$ keV luminosity $L_{\rm X}$ to infer the fraction of IR luminosity (i.e., integrated over $8-1000$ $\mu$m) coming from the central AGN, in a way that is totally independent from SED fitting and avoids all the caveats related to parameter degeneracies (see Sect.~\ref{sec_dust_emission}). In particular, we exploit the correlation by \citet[their Eq. 4]{Mullaney2011:2011MNRAS.414.1082M}:
\begin{equation}\label{Eq_Mullaney_fAGN}
\begin{split}
    & \log\biggl(\frac{L^{\rm AGN}_{\rm IR}}{10^{43}\, {\rm erg\,s^{-1}}}\biggr) = \\
    & = (0.53\pm0.26)+(1.11\pm0.07)\,\log\biggl(\frac{L_{\rm X}}{10^{43}\, {\rm erg\,s^{-1}}}\biggr)
\end{split}
\end{equation}
 to derive the AGN infrared luminosities and the corresponding AGN fraction (i.e., $f_{\rm AGN}^{(1)} = L^{\rm AGN}_{\rm IR}/L^{\rm dust}_{\rm IR}$), that we list in Tab. \ref{tab_mullaney}. Infrared luminosities (Tab. \ref{tab_mullaney}) come from the SED fit with CIGALE, following the approach by \citet[Sect.~\ref{sec_dust_emission}]{Casey2012:2012MNRAS.425.3094C}. 
 
 The correlation by \citet{Mullaney2011:2011MNRAS.414.1082M} is based on a sample of 25 local (i.e., z $<0.1$) AGNs from the Swift-BAT survey, with typical X-ray and IR properties (i.e., $N_{\rm H}$, $L_{2-10\,keV}$ and $L_{\rm IR}$) largely covering the same ranges as those z $\sim2$ AGNs and star-forming galaxies detected in Chandra \citep[e.g., CDF-N and CDF-S; see Fig. 1 in][]{Mullaney2011:2011MNRAS.414.1082M} and Spitzer/Herschel deep surveys (e.g. GOODS; see Sect.~\ref{sec_sample}).  Thus, we can reasonably apply the result from \citet{Mullaney2011:2011MNRAS.414.1082M} to our sample of $z\sim2$ DSFGs.
 It follows that the central AGN contribution to the total infrared light of our X-ray detected DSFGs is negligible (i.e., consistent with $0$ or a few per cent): it attains values $\lesssim10$\% once the 1$\sigma$ scatter (i.e., $\sim0.5$ dex) of Mullaney et al. correlation is considered. A similar result has been found by \citet{Pozzi2012:2012MNRAS.423.1909P} analyzing a sample of $\sim30$ Herschel-selected $z\sim2$ LIRGs: just the $\sim35$\% of the sample show the presence of an AGN at the $3\sigma$ confidence level, but its contribution to the IR emission accounts for only $\sim5$\% of the energy budget.
 
We provide a further investigation on the AGN fraction by referring to the relation between the nuclear 12 $\mu$m luminosity (L$_{12\,\mu m}^{\rm nuc}$) and the intrinsic $2-10$ keV luminosity by \citet[their Eq. 1]{Asmus2015:2015MNRAS.454..766A} found for a local sample but valid also at higher redshift \citep[see also][]{Gandhi2009:2009A&A...502..457G} . We exploited the $2-10$ keV luminosities L$_X$ (Tab. \ref{tab_mullaney}) to derive the expected rest-frame 12 $\mu$m nuclear luminosity of our sources. Comparing the outcome with the corresponding observed luminosity at $\lambda_{\rm rest}=12\,\mu$m, we derive the fraction coming from the nucleus $f_{\rm AGN}^{(2)}$, that we show in the last column of Tab. \ref{tab_mullaney}. The MIR-X ray correlation scatter of 0.34 dex do not allow us to precisely constrain the AGN fraction, but still we can provide a qualitatively estimation of the impact of the central AGN on the observed MIR emission: the fraction of 12 $\mu$m luminosity coming from the nucleus attains to values $\lesssim10\%$ and for the majority of our sources (seven out of nine) it is $\lesssim1\%$. 

We note that the two X-ray non-detected sources (UDF5 and AzTEC.GS22) do not appear also in the supplementary catalog at very low significance. This may indicate either that no (not very powerful) AGN is present or that it is highly obscured, i.e. \textit{Compton-thick}, with N$_H\gtrsim10^{24}$ cm$^{-2}$. However, we can not provide an insight on this issue just basing on the (other) multi-wavelength data at our disposal.

Finally, we cross-checked our results with the outcomes obtained by fitting our DSFG SEDs including the modules from \citet{DraineLi2007:2007ApJ...657..810D} and \citet[]{Fritz2006:2006MNRAS.366..767F} in the  CIGALE routine, while keeping the free parameters fixed to the values found in literature to better reproduce the high-z DSFG emission \citep[see e.g.,][]{Malek2018:2018A&A...620A..50M, Donevski2020:2020A&A...644A.144D}. The resulting AGN fraction $f_{\rm AGN}$ is still smaller than 10\% for almost all the DSFGs of our sample. A similar result has been recently found by \citet{Barrufet2020:2020A&A...641A.129B} by analysing the IR SED of $\sim$ 200 DSFGs in the COSMOS field at $0.7<z_{\rm phot}<5.6$. These evidences provide a further confirmation of our approach in modelling the MIR and FIR emission, as it is discussed in Sect.~\ref{sec_dust_emission}.

\subsection{Radio emission}\label{sec_radio_emission}
\begin{table}
\centering
 \caption{In this table we list in the order: IDs of the sources; the observed frequency in the radio band ($\nu^{\rm obs}$); the corresponding radio fluxes from free-free (F$_{\rm ff}$) and synchrotron emission (F$_{\rm sync}$) by using Equations (\ref{eq_ff}) and (\ref{eq_sync}), for which we adopted a scatter of 0.3 dex; the observed radio flux (F$^{\rm obs}_{\rm radio}$) and their uncertainties. Units are given between square brackets.}\label{tab_radio}
  \begin{tabular}{lcccc}
 \hline
 \bfseries {ID} & $\mathbf{\nu^{obs}}$ & $\mathbf{F_{ff}}$  & $\mathbf{F_{sync}}$ & $\mathbf{F^{obs}_{radio}}$ \\
  & [GHz] & [$\mu$Jy] & [$\mu$Jy] & [$\mu$Jy] \\
 \hline
 UDF1 & 5.25 & $4.0$ & $7.6$ & $9.0\pm0.6$ \\
 UDF3 & 5.25 & $4.5$ & $0.6$ & $12.1\pm0.6$\\
 UDF5 & 5.25 & $2.0$ & $3.8$ & $6.3\pm0.5$\\
 UDF8 & 5.25 & $3.0$ & $5.7$ & $7.2\pm0.5$\\
 UDF10 & 5.25 & $0.7$ & $1.4$ & $ <0.7$\\
 UDF11 & 5.25 & $4.6$ & $8.7$ & $9.3\pm0.7$\\
 UDF13 & 5.25 & $1.4$ & $2.7$ & $4.7\pm0.5$\\
 ALESS067.1 & 1.5 & $9.4$ & $47.3$ & $74\pm7$ \\
   & 0.61 & $10.1$ & $100.3$ & $137\pm15$\\
 AzTEC.GS25 & 1.5 & $6.8$ & $34.1$ & $90\pm6$\\
 AzTEC.GS21 & 1.5 & $8.4$ & $42$ & $44\pm6$\\
 AzTEC.GS22 & 1.5 & $2.4$ & $12.1$ & $35\pm7$\\ 
   \hline
  \end{tabular}
\end{table}

The radio emission in star-forming galaxies can be traced back to two different astrophysical processes: the star formation itself and the accretion of the central SMBH, that can eventually turn into an AGN emitting in the radio band \citep[cfr.][]{Mancuso2017:2017ApJ...842...95M}. 

Radio emission associated with star formation comprises two components: free-free emission coming from HII regions that contain massive, ionizing stars, fully dominating at frequencies $\nu>30$ GHz; synchrotron emission resulting from relativistic electrons accelerated by supernova remnants. In the following we consider both these contributions to provide a rough but realistic estimate of the stellar radio emission for each galaxy of the sample by using the SFR from our SED fitting (see Sect.~\ref{sec_fit_with_cigale}).
We adopt the classical free-free emission calibration with SFR at 33 GHz for a pure hydrogen plasma ($Z_i=1$) with temperature $T=10^4$ K by \citet{Murphy2012:2012ApJ...761...97M} :
\begin{equation}\label{eq_ff}
\begin{split}
    L_{\rm ff} \approx & \, 3.75 \times 10^{26} {\rm erg\, s^{-1} \,Hz^{-1}} \frac{SFR}{M_\odot\,{\rm yr}^{-1}} \biggl(\frac{T}{10^4 \,{\rm K}}\biggr)^{0.3} \times \\
    & \times g(\nu,\,T) \, \exp\biggl(-\frac{h\nu}{kT}\biggr)
\end{split}
\end{equation}
where $g(\nu,\,T)$ is the Gaunt factor: approximated according to \citet{Draine2011:2011piim.book.....D}. Synchrotron calibration with SFR is a bit controversial since it involves complex and poorly understood processes, such as the production and escaping rates of relativistic electrons and the magnetic field strength. Here we exploit the calibration proposed by \citet{Murphy2011:2011ApJ...737...67M, Murphy2012:2012ApJ...761...97M}, but see also the review by \citet{KennicuttEvans2012:2012ARA&A..50..531K}. 
Thus, the synchrotron luminosity ascribed to star formation can be written as follows:
\begin{equation}\label{eq_sync}
\begin{split}
    L_{\rm sync}\approx & \,1.9\times10^{28}{\rm erg\,s^{-1}\,Hz^{-1}} \frac{SFR}{\rm M_\odot\,yr^{-1}}\Bigl(\frac{\nu}{{\rm GHz}}\Bigr)^{-\alpha_{\rm sync}} \times\\
    & \times \biggl(1+\Bigl(\frac{\nu}{20\,\rm GHz}\Bigr)^{0.5}\biggr)^{-1}\times\frac{1 -\exp(-\tau_{\rm sync}(\nu))}{\tau_{\rm sync}(\nu)} 
\end{split}
\end{equation}
where $\alpha_{\rm sync}\sim0.75$ is the spectral index found for high-z DSFGs \citep[e.g.,][]{Condon1992:1992ARA&A..30..575C, Ibar2009:2009MNRAS.397..281I, Ibar2010:2010MNRAS.401L..53I, Thomson2014:2014MNRAS.442..577T}, the term $(1+\nu^{0.5})^{-1}$ renders spectral aging effects \citep[see][]{BandayWolfendale1991:1991MNRAS.248..705B}, and the latter factor takes into account synchrotron self-absorption in terms of the plasma optical depth \citep[e.g.,][]{Kellermann1966:1966ApJ...146..621K,Tingay2003:2003AJ....126..723T}.

Then, we compare these predictions with the observed radio fluxes (Tab. \ref{tab_radio}) in order to get some hints on the presence (or not) of a central AGN. For every source we find the radio emission to be consistent with galaxy star formation (see Fig.~\ref{panel}): radio fluxes lie within the scatter of free-free plus synchrotron radio emission, represented by the yellow shaded area. It is worth noticing, though, that this evidence does not exclude the presence of a central AGN, whose radio emission could be simply too low to emerge from the stellar one. We pinpoint three possible scenarios: the galaxy does not host an AGN; the galaxy host an accreting central SMBH but it does not contribute to the observed emission in the radio band; an AGN is present but it is radio silent or radio quiet. In this respect we provide a further analysis in Section \ref{sec_results_discussion}. 

\subsection{Multi-wavelength source sizes}\label{sec_multiwavelength_sizes}

\begin{table*}
\centering
 \caption{Circularized radii for the multi-wavelength counterparts of our (sub-)millimetre sources (when resolved in the corresponding map). Source ID in column 1. $c$ (third column) is the number of proper kpc at the redshift of the source; H-band circularized radii ($\rm {r_H}$) in the fourth column are derived from the effective S\'{e}rsic half-light semi-axis by \citet{vanderWel2012:2012ApJS..203...24V} and $n$ is the corresponding S\'{e}rsic index of the light radial profile (fifth column); ALMA and VLA circularized radii ($ {\rm r_{\rm ALMA}}$ and ${\rm r_{\rm VLA}}$ respectively are listed in the last two columns. Flag $^\blacklozenge$: the fit to the source is suspicious \citep[][]{vanderWel2012:2012ApJS..203...24V}}\label{tab_sizes}
\begin{tabular}{lccccc}
 $\mathbf{ID}$  & \bfseries {\textit{c}}  & \bfseries {\textit{n}}$\mathbf{_{HST}}$ &$\mathbf{r_{HST}}$ & $\mathbf{r_{ALMA}}$  & $\mathbf{r_{VLA}}$ \\
  &  [kpc/'']& & [kpc] & [kpc] & [kpc] \\
 \hline
  UDF1 & 8.121 & $7.2\pm0.4$ & $0.57\pm0.02$ & $1.46\pm0.06$ &   $2.8\pm0.1$\\ 
  UDF3 & 8.224 & $0.81\pm0.02^\blacklozenge$ & $1.59\pm0.01$ & $1.85\pm0.06$ &  $1.48\pm0.03$\\
  UDF5 & 8.632 & $0.71\pm0.01$ & $2.29\pm0.01$ & $1.8\pm0.1$  & $2.4\pm0.1$ \\ 
  UDF8 & 8.647 & $3.0\pm0.02$ & $5.68\pm0.01$ & $4.3\pm0.5$ & $2.1\pm0.1$  \\  
  UDF10 & 8.508 & $1.23\pm0.02$ & $1.96\pm0.01$ & $-$ & $-$ \\
  UDF11 & 8.551 & $1.41\pm0.01$ & $4.50\pm0.01$ & $4.3\pm0.6$ & $3.4\pm0.3$ \\
  &   & $-$ &  &   $-$    &   $0.65\pm0.04$ \\
  UDF13 & 8.256 & $1.86\pm0.03$ & $1.17\pm0.01$  & $2.6\pm0.4$ & $2.1\pm0.2$\\
  ALESS067.1 & 8.489 & $7.9\pm0.7$ & $6.5\pm0.2$ & $1.0\pm0.1$ & $-$\\
  AzTEC.GS25 & 8.390 & $3.5\pm0.5$ & $1.8\pm0.1$ & $0.9\pm0.04$ & $-$ \\
  AzTEC.GS21 & 8.586 & $1.50\pm 0.06$ & $3.66\pm0.05$ & $-$ & $-$ \\
  AzTEC.GS22 & 8.624 & $0.22\pm0.03^\blacklozenge$ & $3.2\pm0.1$ & $-$ & $-$\\
  \hline
  \end{tabular}
 \end{table*}
 
 Comparing multi-wavelength sizes and centroid positions is crucial to infer the evolutionary phase of high-z DSFGs \citep[e.g.,][]{Lapi2018a:2018ApJ...857...22L} and differences in ISM conditions \citep[e.g.,][]{Elbaz2018:2018A&A...616A.110E,Donevski2020:2020A&A...644A.144D}.  To homogenize the information over the whole sample, we derive the linear circularized multi-wavelength radius by using the following expression:
\begin{equation}\label{eq_sizes}
 r_{circ}\,[{\rm kpc}] = {\rm R} \,[{\rm arcsec}]\,\, \sqrt{\rm q} \,\,c\,[{\rm kpc}/{\rm arcsec}]
\end{equation}
when the corresponding source angular size R (i.e. half-light semi-major axis R$_{1/2}$ or effective radius R$_{e}$, the latter in case of a S\'{e}rsic profile fit of the source light profile) is available in the literature.
In Eq. \eqref{eq_sizes}, q is the projected axis ratio and $c$ is the angular-to-linear conversion factor, which depends on redshift and cosmology (see Tab. \ref{tab_sizes}). The HST H-band circularized radii are derived from the angular sizes measured by \citet{vanderWel2012:2012ApJS..203...24V}, who performed a S\'{e}rsic fit of the resolved sources in the CANDELS HST survey. The ALMA and VLA circularised radii of the HUDF sources (UDF1  $-$ UDF13 ) are derived from the corresponding angular sizes measured in the source deconvolved FWHMs images by \citet{Rujopakarn2016:2016ApJ...833...12R}, who performed a two-dimensional (2D) elliptical Gaussian fitting allowing for multiple components. This is the case of  UDF11 radio counterpart  (see Tab. \ref{tab_sizes}) showing two main gaussian components: the first is centered on the ALMA source while the second extends around it \citep[cfr. Fig 1b in][]{Rujopakarn2016:2016ApJ...833...12R}. As to the AzTEC and LABOCA sources (ALESS067.1  $-$ AzTEC.GS22), (sub-)millimetre and radio sizes are not available since they are non resolved in the corresponding maps. Just for two sources (ALESS067.1, AzTEC.GS22) we found the angular size (i.e., S\'{e}rsic effective radius R$_{\rm e}$) of their ALMA counterparts in the DANCING ALMA catalog by \citet{Fujimoto2017:2017ApJ...850...83F}, that we exploited to derive the circularized radii.
 
\section{Discussion}\label{sec_results_discussion}

In this Section we discuss our results from two diverse perspectives. First, we provide a general overview on the sample, characterizing our galaxies by exploiting the main results from SED fitting and providing a further analysis on their multi-wavelength emission, that we have extensively described in Sects. \ref{sec_fit_with_cigale} and \ref{sec_add_info}. Then, we place the sources in the broader context of galaxy evolution, both comparing with the most popular diagnostic plots (that are empirically derived, such as galaxy main-sequence; dust mass and gas mass vs. stellar mass; gas metallicity relation), and predictions from theory \citep[e.g.,][]{Pantoni2019:2019ApJ...880..129P}. To this aim, we mainly refer to the \textit{in-situ} galaxy-BH co-evolution scenario \citep[see e.g.,][]{Mancuso2016a:2016ApJ...823..128M, Mancuso2016b:2016ApJ...833..152M, Mancuso2017:2017ApJ...842...95M, Lapi2018a:2018ApJ...857...22L}.

\begin{table}
 \centering
 \caption{Median, first  and third quartiles of the following quantities (in the order): redshift ($z$), age of the burst ($\tau_\star$); SFR; stellar mass (M$_\star$); ISM attenuation spectral index ($\delta_{\rm ISM}$); IR luminosity (L$_{\rm IR}$); dust mass (M$_{\rm dust}$); gas mass (M$_{\rm gas}$: total and molecular); depletion time ($\tau_{\rm depl}$); AGN fraction in the IR domain ($f_{\rm AGN}$); ALMA, HST and VLA sizes ($r_{\rm ALMA}$, $r_{\rm HST}$ and $r_{\rm VLA}$). The ($\times2$) in $M_{\rm dust}$ has to be considered to apply the correction by \citet{Magdis2012;2012ApJ...760....6M} (see Sect.~\ref{sec_dust_mass}). }\label{tab_sample_median_value}
  \begin{tabular}{llccc}
  \hline
   & & \textbf{Median} & $\mathbf{1^{st}}$ \textbf{quartile} & $\mathbf{3^{rd}}$ \textbf{quartile}\\
 \hline
   z & & 2.086 & 1.794 & 2.497\\
   SFR & [M$_\odot$ yr$^{-1}$] & 241 & 91 & 401 \\
   $\tau_\star$ & [Myr] & 746 & 334 & 917 \\
   M$_{\star}$ &  [$10^{10}$ M$_\odot$] & 6.5 & 5.7 & 9 \\
   $\delta_{\rm ISM}$ & & -0.5 & -0.1 & -0.7\\
   $f_{\rm att}$ & [$10^{-4}$] & 2.5 & 1 & 3\\
   L$_{\rm IR}$ & [$10^{12}$ L$_\odot$] & 2.2 & 1.01 & 3.9 \\
   M$_{\rm dust}$ & [$10^{8}$ M$_\odot$] & ($2\times$)2.3 & ($2\times$)0.95 & ($2\times$)4.8 \\
   M$_{\rm gas,\,tot}$ & [$10^{10}$ M$_\odot$] & 6.3 & 3.2 & 10.0\\
   M$_{\rm gas,\,mol}$ & [$10^{10}$ M$_\odot$] & 3.2 & 0.6 & 5.0\\
   $\tau_{\rm depl}$ & [Myr] & 205 & 143 & 395\\ 
   f$_{\rm AGN}$ & [\%] & 0.8 & 0.4 & 1\\ 
   r$_{\rm ALMA}$ & [kpc] & 1.8 & 1.2 & 3.5 \\
   r$_{\rm HST}$ & [kpc] & 2.3 & 1.6 & 4.5 \\
   r$_{\rm VLA}$ & [kpc] & 2.25 & 2.1 & 2.8 \\
  \hline
  \end{tabular}
\end{table}
In Tab. \ref{tab_sample_median_value} we list the median physical properties of the sample and the corresponding first and third quartiles. 
Our 11 galaxies are young (median $\tau_\star\sim0.7$ Gyr) and forming stars at high rates, of the order of hundreds M$_\odot$ yr$^{-1}$, leading to stellar masses of $\sim10^{10}-10^{11}$ M$_\odot$. 
This very intense star formation activity is typically observed in the very central regions of the galaxies. From high-resolution imaging (when available) we found our galaxies to be compact, both in the FIR/mm (median r$_{\rm ALMA} \sim 1.8$ kpc), in the optical (median r$_{\rm HST} \sim 2.3$ kpc) and in the radio band (median r$_{\rm VLA} \sim 2.25$ kpc).
Gas-rich (median total M$_{\rm gas,\,tot}\sim6\times10^{10}$ M$_\odot$; median molecular M$_{\rm gas,\,mol}\sim3-4\times10^{10}$ M$_\odot$) and characterized by depletion timescales of a (few) hundred(s) of Myr (median $\tau_{\rm depl}\sim100$ Myr), our objects are the typical high-z (sub-)millimetre star-forming galaxies whose detections have been constantly growing since the advent of ALMA \citep[e.g.,][for a review]{Tadaki2015:2015ApJ...811L...3T,Massardi2018:2018A&A...610A..53M, Talia2018:2018MNRAS.476.3956T, HodgeDaCunha2020:2020arXiv200400934H}. 
They have high IR luminosities ($\sim$ a few $10^{12}$ L$_\odot$), comparable to the typical values of the local ULIRGs and high-z DSFGs \citep[see the review by][]{CaseyNarayananCooray2014:2014PhR...541...45C}, revealing a large interstellar dust content (median M$_{\rm dust}\sim5\times10^{8}$ M$_\odot$). AGN fraction contributing to IR emission is negligible (f$_{\rm AGN}=1$\% lies in the $75^{th}$ percentile).
Similar values are found in literature \citep[e.g.,][]{Elbaz2018:2018A&A...616A.110E, Barrufet2020:2020A&A...641A.129B}. As to the ISM attenuation law, it results shallower (median $\delta_{\rm ISM}\sim - 0.5$) than the standard \citet{Calzetti2000:2000ApJ...533..682C}, constrained in local star-forming galaxies. 
Our results, mainly derived from the SED fitting, basically reflect the selection we have performed on the FIR/mm catalogs in the GOODS-S field to build our sample (see Sect.~\ref{sec_sample}).  

\begin{figure}
  \centering
\includegraphics[width=\columnwidth]{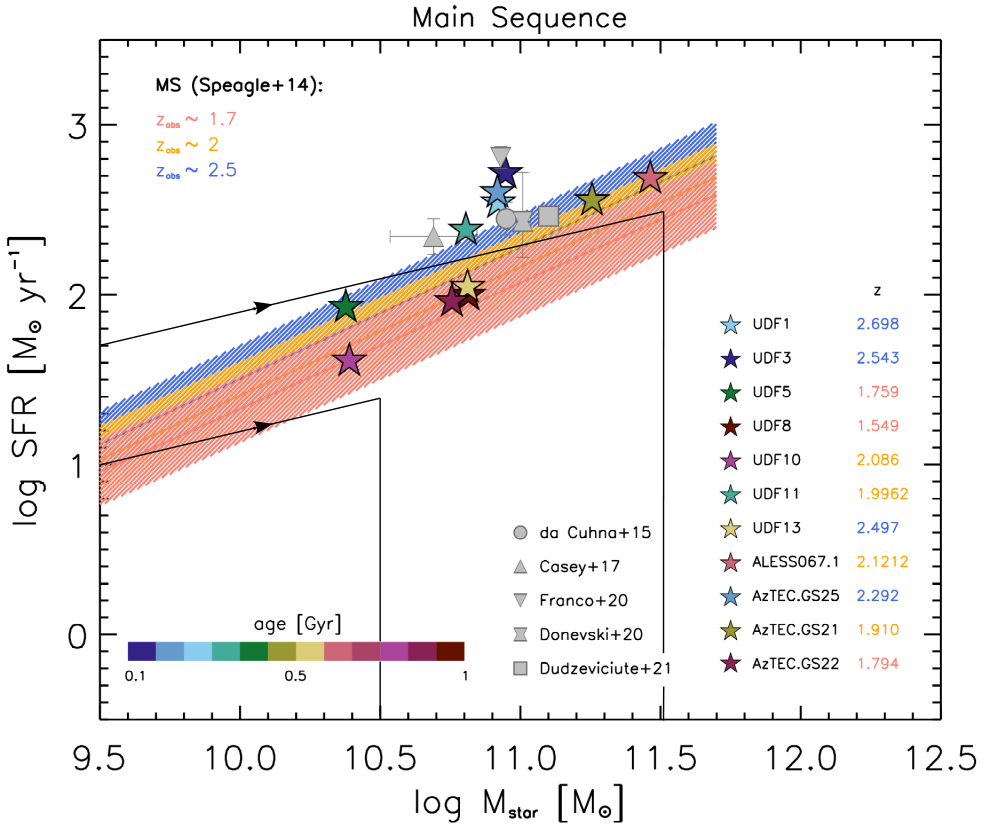}
\caption{Empirical galaxy main-sequence of star-forming galaxies by \citet{Speagle2014:2014ApJS..214...15S} at different redshifts: $z\sim1.5$ in red, $z\sim2$ in orange, $z\sim3.5$ in blue, with its 1$\sigma$ scatter ($\sim 0.2$ dex). Over-plotted stars stand for the galaxies of this work, considering the values of SFR and M$_\star$ derived from our SED fitting. Symbols are color-coded by galaxy age (i.e., $\tau_\star$). Error bars are compatible with symbol size.  Galaxy redshift  is indicated in the legend next to galaxy ID and color-coded by the corresponding redshift bin. Grey symbols represent the median values obtained by some other existing samples of high-z DSFGs, as specified in legend. For reference, we plot the evolutionary tracks predicted by the \textit{in-situ} co-evolution scenario \citep[black solid lines;][]{Mancuso2016b:2016ApJ...833..152M}.}
\label{fig_MS_evo_lines}
\end{figure}

In Fig.~\ref{fig_MS_evo_lines} we place the galaxies of our sample on the ${\rm M}_\star-$ SFR plane. SFRs, stellar masses and galaxy ages are derived from SED fitting, as described in Sect.~\ref{sec_fit_with_cigale}. In the last decades, the majority of star-forming galaxies, from local to high-z Universe (at least out to $z\sim4$), have been found to follow an empirical relation between stellar mass and SFR, the galaxy main-sequence \citep[e.g.,][]{Daddi2007:2007ApJ...670..156D, Noeske2007:2007ApJ...660L..43N}. In Fig.~\ref{fig_MS_evo_lines} we compare the position of our DSFGs with the empirical determination of galaxy main-sequence by \citet{Speagle2014:2014ApJS..214...15S}, at redshifts 1.7, 2, and 2.5 (color-coded), spanning the redshift range of the sample ($z_{\rm spec}\sim1.5-3$). We note that our sources lie in correspondence or just above the relation at the corresponding redshift.

Interpreting the diverse \textit{loci} of star-forming galaxies onto the ${\rm M}_\star-$ SFR plane is very controversial, since it strongly depends on the scenario accounted for describing galaxy evolution. The debate concerning the main drivers of DSFGs formation and evolution through cosmic time is still object of intense discussions and we are currently not able to comprehensively solve the issue. 
To interpret our outcomes we refer to the \textit{in-situ} galaxy-BH co-evolution scenario \citep[see e.g.,][]{Eke2000:2000MNRAS.315L..18E, Fall2002:2002ASPC..275..389F,Romanowsky2012:2012ApJS..203...17R, Mancuso2016a:2016ApJ...823..128M, Mancuso2016b:2016ApJ...833..152M, Mancuso2017:2017ApJ...842...95M, Shi2017:2017ApJ...843..105S, Lapi2018a:2018ApJ...857...22L}. Our theoretical interpretation (possibly) does not constitute the only consistent description of our results \citep[other approaches can be found in literature, e.g.,][]{Dekel2006:2006MNRAS.368....2D, Hopkins2006:2006ApJ...652..864H, Calura2017:2017MNRAS.465...54C, Eales2017:2017MNRAS.465.3125E}. In the following we want just to provide a possible and self-consistent interpretation that encompasses galaxy broad-band and spectroscopic emission, multi-wavelength sizes and galaxy co-evolution with the central SMBH. To get more details on the main steps characterizing the formation and evolution of massive high-z IR luminous galaxies, as predicted by the \textit{in-situ} scenario, we refer the reader to \citet{Lapi2018a:2018ApJ...857...22L}. A schematic view of galaxy typical SFH and BH Accretion History (BHAH) is available in \citet[their Fig. 2]{Mancuso2017:2017ApJ...842...95M}. 

In Fig.~\ref{fig_MS_evo_lines} we show a pair of tracks which represent the path onto ${\rm M}_\star-$ SFR plane followed by a high-z DSFG during its evolution, as predicted by the \textit{in-situ} galaxy-BH co-evolution scenario (solid black lines). Time flow is indicated by the black arrows. The starting point of galaxy evolutionary track is determined by galaxy SFR when star formation ignites. During the early phase, star formation proceeds at an almost constant rate (i.e., SFR $\propto \tau^{1/2}$ and M$_\star\propto\tau^{2/3}$, which implies SFR $\propto$ M$_\star^{1/3}$) and, increasing its stellar mass, the galaxy approaches the main-sequence of star-forming galaxies \citep[][see their Eq. 7]{Mancuso2016b:2016ApJ...833..152M}. Galaxy main-sequence emerges as a statistical \textit{locus} in ${\rm M}_\star-$ SFR plane, where it is more probable to find star-forming galaxies because they spend in its vicinity most of their lifetime. 
Following the \textit{in-situ} scenario, the less abundant population of star-forming galaxy that has been found to lie above the main-sequence \citep[traditionally referred as \textit{starbursts}; e.g.,][]{Rodighiero2011:2011ApJ...739L..40R} is constituted by young galaxies that have still to accumulate most of their stellar mass. As soon as AGN feedback removes the fuel of star formation (effective for galaxies with M$_\star>10^{10}$ M$_\odot$), galaxy SFR is abruptly reduced and the object moves below the main-sequence, becoming a red and dead galaxy. Stars in Fig.~\ref{fig_MS_evo_lines} represent the objects of our sample and are color-coded by galaxy age (i.e., $\tau_\star$), as obtained from SED fitting. Younger (i.e., bluer) objects are found to lie to the upper-left side of the main-sequence at the corresponding redshift, while the elder (i.e., redder) are found to lie in correspondence of it, as predicted by the \textit{in-situ} scenario. All in all, these galaxies are almost main-sequence objects and we expect for them to find some signatures of obscured and/or accreting AGN, and possibly some evidences of its activity (i.e., outflows/winds). However, its contribution to the galaxy IR emission is negligible: the median value for our sample attains less than $1$\%.

We note that the median values of stellar mass and SFR of our sample (i.e., median M$_\star\sim6.5\times10^{10}$ M$_\odot$ and median SFR$\sim241$ M$\odot$ yr$^{-1}$) are consistent (i.e., lie in the same area of M$_\star-$ SFR plane) with the median values found in the most recent studies on high-z DSFGs exploiting SED fitting \citep[e.g.][grey symbols in Fig. \ref{fig_MS_evo_lines}]{daCunha2015:2015ApJ...806..110D, Casey2017:2017ApJ...840..101C, Franco2020:2020A&A...643A..30F, Donevski2020:2020A&A...644A.144D, Dudzeviciute2021:2021MNRAS.500..942D} spanning the photometric redshift range $0.5<z<5$. These results refer to samples of different sizes, including large statistically significant samples of DSFGs selected in the FIR-millimetre domain \citep[][]{Donevski2020:2020A&A...644A.144D, Dudzeviciute2021:2021MNRAS.500..942D}, and smaller samples of a few tenths of objects, with different selection criteria.

The typical compactness of DSFGs revealed by ALMA in the recent past years \citep[typical FIR radii $\sim1$ to a few kpc; e.g.,][]{Negrello2014:2014MNRAS.440.1999N, Riechers2014:2014ApJ...796...84R, Ikarashi2015:2015ApJ...810..133I, Dye2015:2015MNRAS.452.2258D, Simpson2015:2015ApJ...807..128S, Tadaki2015:2015ApJ...811L...3T, Harrison2016:2016MNRAS.457L.122H, Scoville2016:2016ApJ...820...83S, Dunlop2017:2017MNRAS.466..861D, Massardi2018:2018A&A...610A..53M, Talia2018:2018MNRAS.476.3956T}, together with the evidence from high-resolution imaging that the most intense events of star formation occur in a few collapsing clumps located in this very central region of the galaxy \citep[e.g.,][]{Hodge2019:2019ApJ...876..130H, Tadaki2019:2019ApJ...876....1T}, provide an observational confirmation of the \textit{in-situ} scenario predictions for the evolution of massive high-z DSFGs. The whole multi-band size evolution predicted by the \textit{in-situ} scenario \citep[see][]{Lapi2018a:2018ApJ...857...22L} is consistent with the median optical and FIR size we found for our sample (i.e., r$_{\rm HST}\sim2.3\,{\rm kpc}>{\rm r}_{\rm ALMA}\sim 1.8$ kpc; see Tab. \ref{tab_sample_median_value}) and it has been recently confirmed by \citet{Tadaki2020:2020ApJ...901...74T}. Most of the sources show an optical isolated morphology, while 4 galaxies (UDF11, ALESS067.1, AzTEC.GS21, AzTEC.GS22; Pantoni et al. in preparation) have more complex (i.e., clumpy) morphologies, possibly indicating the presence of minor companions - that can enhance the formation of stars and prolong the star formation in the dominant galaxy by refuelling it with gas - or just being a signature of the ongoing dusty star formation. The picture is even more uncertain, since no photometric redshift has been measured for the companion candidates.

\begin{figure}
  \centering
\includegraphics[width=\columnwidth]{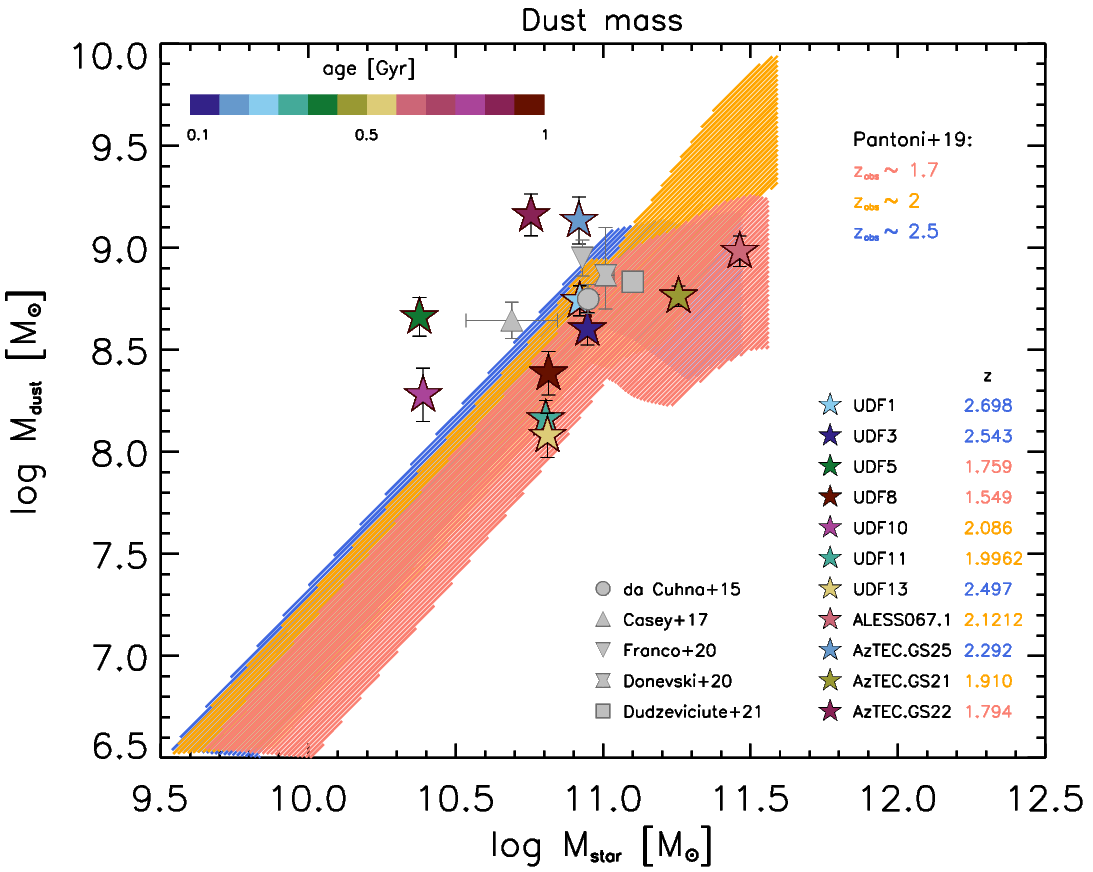}
\caption{Statistical relationship between dust mass M$_{\rm dust}$ and stellar mass M$_{\star}$ for the high-z star-forming progenitors of ETGs by \citet{Pantoni2019:2019ApJ...880..129P} at three different observational redshifts: $z\sim1.5$ (red), $z\sim2$ (orange), $z\sim2.5$ (blue), with its 1$\sigma$ scatter (shaded area). Stars represent the outcomes for the 11 DSFGs of our sample derived from SED fitting as explained in Sect.~\ref{sec_dust_mass} and corrected following \citet{Magdis2012;2012ApJ...760....6M}. Symbols are color-coded by galaxy age (i.e., $\tau_\star$). Galaxy redshift is indicated in the legend next to galaxy ID and color-coded by the corresponding redshift bin. Grey symbols represent the median values obtained by some other existing samples of high-z DSFGs, as specified in legend.}\label{fig_Mdust}
\end{figure}

\begin{figure}
  \centering
\includegraphics[width=\columnwidth]{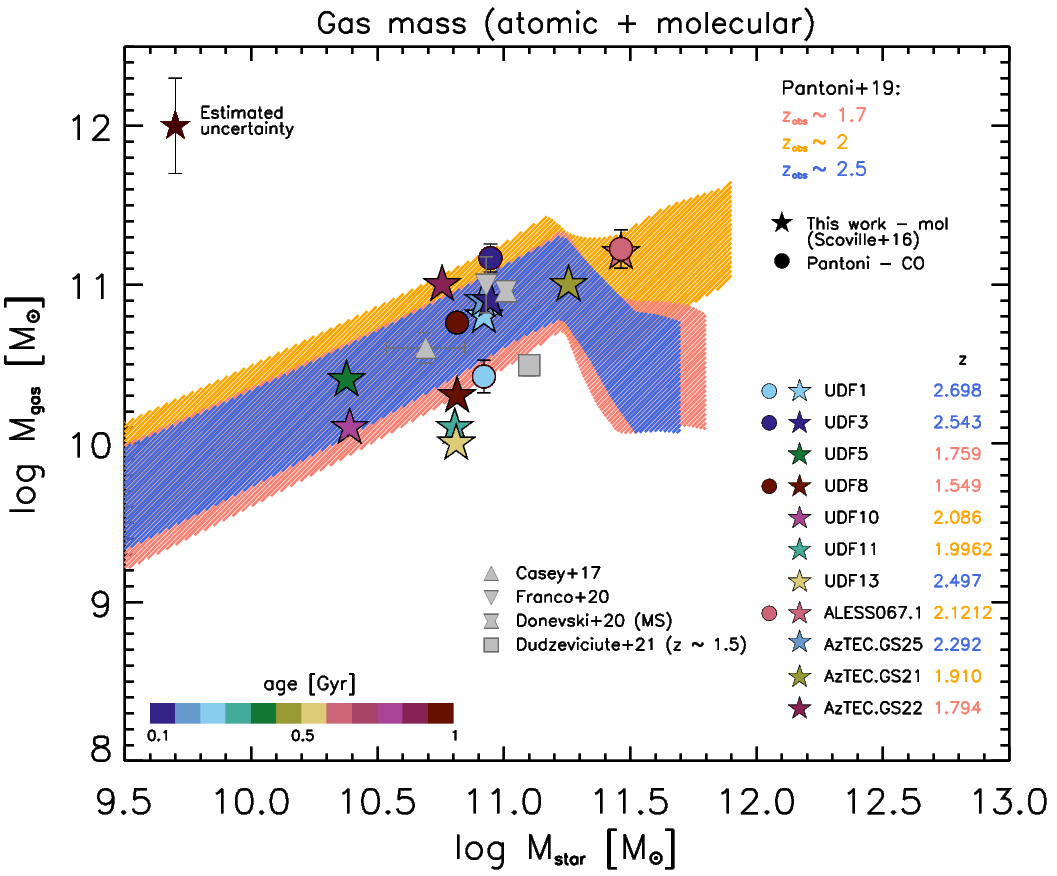}
\caption{Statistical relationship between gas mass M$_{\rm gas}$ and stellar mass M$_{\star}$ for the high-z star-forming progenitors of ETGs by \citet{Pantoni2019:2019ApJ...880..129P} at three different observational redshifts: $z\sim1.5$ (red), $z\sim2$ (orange), $z\sim2.5$ (blue), with its 1$\sigma$ scatter (shaded area). Stars represent the outcomes for the 11 DSFGs of our sample derived from SED fitting as explained in Sect.~\ref{sec_gas_mass}, following the approach by \citet{Scoville2016:2016ApJ...820...83S}. The estimated uncertainty ($\sim0.3$ dex) is consistent with the 1$\sigma$ scatter of the relation and is shown in the left top corner of the plot. Circles stand for the H$_2$ masses estimated from $J>1$ CO lines by and by Pantoni et al. (in preparation). Symbols are color-coded by galaxy age (i.e., $\tau_\star$). Galaxy redshift is indicated in the legend next to galaxy ID and color-coded by the corresponding redshift bin. Grey symbols represent the median values obtained by some other existing samples of high-z DSFGs, as specified in legend.}\label{fig_Mgas_scoville}
\end{figure}

\begin{figure}
  \centering
\includegraphics[width=\columnwidth]{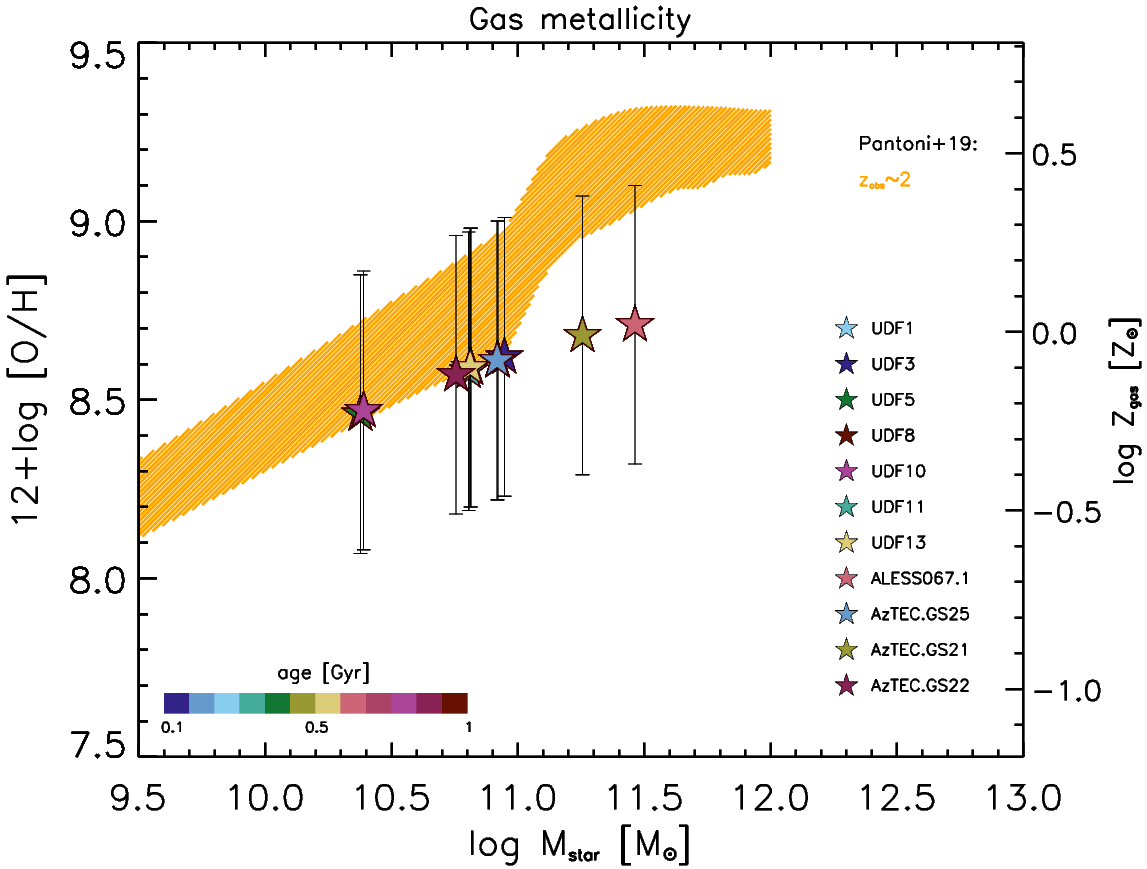}
\caption{Statistical relationship between gas metallicity Z$_{\rm gas}$ and stellar mass M$_{\star}$ for high-z star-forming progenitors of ETGs, by \citet{Pantoni2019:2019ApJ...880..129P} at $z\sim2$ (orange), with its 1$\sigma$ scatter (shaded area). Stars represent our outcomes for the 11 DSFGs of our sample derived as explained in Sect.~\ref{sec_gas_mass}. Symbols are color-coded by galaxy age (i.e., $\tau_\star$).}\label{fig_gas_metallicity_elbaz}
\end{figure}

Figs. \ref{fig_Mdust}, \ref{fig_Mgas_scoville} and \ref{fig_gas_metallicity_elbaz} show the statistical relationships found by \citet{Pantoni2019:2019ApJ...880..129P}, with their 1$\sigma$ scatter, at redshifts 1.7, 2, and 2.5 (color-coded). In this work, the authors present a new set of analytic solutions aimed at self-consistently describing the spatially-averaged time evolution of the gas, dust, stellar and metals content in star-forming galaxies, focusing on the high-z counterparts of local ETGs. The main statistical relationships are derived after setting the main parameters to follow the framework defined by the \textit{in-situ} galaxy-BH co-evolution scenario. Stars represent our 11 DSFGs and are color coded by galaxy age (i.e., $\tau_\star$). We note that they match the model prediction within 2$\sigma$, again witnessing that the main drivers of the evolution of our 11 DSFGs can be traced back mostly to in-situ processes. The superimposed grey symbols represent the median values found by \citet{daCunha2015:2015ApJ...806..110D, Casey2017:2017ApJ...840..101C, Franco2020:2020A&A...643A..30F, Donevski2020:2020A&A...644A.144D, Dudzeviciute2021:2021MNRAS.500..942D} as specified in the legend. The corresponding outcomes, found also for statistical samples of DSFGs in the photometric redshift range $0.5<z_{\rm phot}<5$ \citep[see][]{Donevski2020:2020A&A...644A.144D, Dudzeviciute2021:2021MNRAS.500..942D}, are in agreement with both the model predictions by \citet[]{Pantoni2019:2019ApJ...880..129P} and the corresponding median values of our sample, i.e. M$_{\rm dust}\sim4.6\times10^8$ M$_\odot$ and M$_{\rm gas, \, mol}\sim3.2\times10^{10}$ M$_\odot$ (crf. Tab. \ref{tab_sample_median_value}). It follows that our approach and selection criteria (see Sect. \ref{sec_sample}) do not introduce any substantial bias and may be applied to statistical samples of spectroscopically-confirmed DSFGs, as soon as they will be available (see e.g., the ongoing z-GAL NOEMA Large Program\ref{footnote1}; PIs: P. Cox; T. Bakx; H. Dannerbauer). In the following we briefly comment on our outcomes, going into more details. 

We derived the dust masses of our DSFGs, shown in Fig. \ref{fig_Mdust}, as described in Sect. \ref{sec_dust_mass}. Then, we corrected the outcomes by a factor 2 following \citet{Magdis2012;2012ApJ...760....6M}. All the galaxies have a very high content of interstellar dust (M$_{\rm dust}>10^8$ M$_\odot$), that is almost consistent with the relations by \citet{Pantoni2019:2019ApJ...880..129P}, who predict a very rapid ($\sim10^7-10^8$ yr) pollution of the ambient by dust and metals. We note that the interstellar dust content does not show any significant trend with galaxy age (i.e. the burst duration, $\tau_\star$). Four galaxies (UDF5, UDF10, AzTEC.GS25, AzTEC.GS22) are outliers (but still consistent with the statistical relation within 2$\sigma$): although they are very close to galaxy main-sequence at the corresponding redshift (crf. Fig. \ref{fig_MS_evo_lines}), they show a dust-to-stellar mass ratio very similar to that of ALMA starbursts in the sample studied by \citet[]{Donevski2020:2020A&A...644A.144D}, i.e. M$_{\rm dust}/$M$_\star\gtrsim0.01$. It could indicate that they are experiencing a quicker growth of dust in their ISM (on timescales shorter than $10^8$ yr), or that they are characterized by much longer dust destruction timescales, preserving larger grains longer \citep[see][their Sect. 4]{Donevski2020:2020A&A...644A.144D}.

In Fig.~\ref{fig_Mgas_scoville} we compare the molecular gas mass estimates for our 11 DSFGs (stars and circles, color coded by galaxy age, i.e., $\tau_\star$) with the predictions by \citet[colored shaded area, including their $1\sigma$ scatter, for 3 redshift bins spanning the sample redshift range]{Pantoni2019:2019ApJ...880..129P} and the median values recently found by some studies of high-z DSFGs exploiting SED fitting \citep[i.e.,][grey symbols]{Casey2017:2017ApJ...840..101C,Franco2020:2020A&A...643A..30F,Donevski2020:2020A&A...644A.144D,Dudzeviciute2021:2021MNRAS.500..942D}. We find them in great accordance, within the errors. No significant trend emerges, neither in redshift or galaxy age. Stars represent our molecular gas mass estimates that we derived from dust FIR continuum following the approach by \citet{Scoville2016:2016ApJ...820...83S}, as described in Sect.~\ref{sec_gas_mass}. Error bars, shown in the top left corner of Fig.~\ref{fig_Mgas_scoville}, are comparable to the 1$\sigma$ scatter of the statistical relationships by \citet{Pantoni2019:2019ApJ...880..129P}. Even if the latter predicts the evolution of the total gas mass, i.e. $HI+H_2$, with galaxy stellar mass (at a given redshift), comparing the two is still valid since we expect the molecular phase to be definitely dominant for these objects. Circles represent the estimates for the hydrogen molecular mass derived from $J>1$ CO lines (Pantoni et al. in preparation; see the discussion Sect.~\ref{sec_gas_mass}), that are listed in Tab. \ref{tab_decarli} and available for four galaxies, i.e. UDF1, UDF3, UDF8 and ALESS067.1. We derived the $H_2$ masses by assuming an $\alpha_{\rm CO}=3.6$ M$_\odot$(K km s$^{-1}$ pc$^2$ )$^{-1}$ (see the discussion in Sect.~\ref{sec_gas_mass}), consistently with the approach followed by \citet{Scoville2016:2016ApJ...820...83S}. The outcomes are almost in agreement within the error bars (the comparison is sensible as $H_2$ constitutes the large majority of the molecular gas content in galaxies: even when corrected for Helium mass by a factor 1.36, the estimations are still consistent). However, some differences can be traced back to the fact that dust and CO emission often sample diverse (more compact/extended) galaxy regions, respectively \citep[see e.g.,][]{Carilli2013:2013ARA&A..51..105C, Scoville2014:2014ApJ...783...84S,Scoville2016:2016ApJ...820...83S,Scoville2017:2017ApJ...837..150S,Elbaz2018:2018A&A...616A.110E}.

All the galaxies have almost solar metallicities (even if error bars are huge; see Fig.~\ref{fig_gas_metallicity_elbaz}), and do not show any significant trend with galaxy age. We compare the gas metallicity just with the statistical relation by \citet[]{Pantoni2019:2019ApJ...880..129P} at $z\sim2$, since it does not show a significant evolution in redshift and the gas metallicity of our DSFGs is not strictly constrained due to the big uncertainties, i.e. $\sim0.4$ dex.

The large content of cold gas ($10^{10}\lesssim{\rm M}_{\rm gas,mol}/{\rm M}_\odot\lesssim10^{11}$), interstellar dust (M$_{\rm dust}>10^8$ M$_\odot$) and stars ($3\times10^{10}\lesssim{\rm M}_\star/{\rm M}_\odot\lesssim3\times10^{11}$), associated with relatively short depletion timescales ($\sim200$ Myr) and compact sizes in the ALMA continuum (r$_{\rm ALMA}\lesssim2$ kpc), suggest that our 11 DSFGs are the high-z star-forming progenitors of (compact) massive elliptical galaxies, caught in the \textit{compaction phase}, i.e. a phase characterized by clump/gas migration toward the galaxy center, where the intense dust-enshrouded star formation takes place and most of the stellar mass is accumulated \citep[see][]{Lapi2018a:2018ApJ...857...22L}. This statement is  furthermore confirmed by the detection of an X-ray emitting AGN for the majority of DSFGs in our sample (Sect. \ref{sec_Xray_analysis}), that does not emerge in the radio domain. Indeed, while the star formation ignites in the host galaxy, the \textit{in-situ} scenario predicts the growth of the central BH to occur at mild super-Eddington rates so that rotational energy cannot be easily funnelled into jets to power radio emission: the AGN is expected to be radio silent and to shine as an X-ray source. This is possibly the case of the objects in our sample that are located to the left-upper side of galaxy main-sequence (crf. Fig. \ref{fig_MS_evo_lines}). Then, following the \textit{in-situ} scenario, we expect AGN X-ray luminosity to overwhelm that associated with star formation, becoming clearly detectable at luminosities $L_X\gtrsim10^{42}$ erg s$^{-1}$ (crf. Fig. \ref{fig_ranalli_age}: most of our DSFGs have $L_X>10^{42}$). AGN power progressively increases to values similar to or even exceeding that of star formation in the host galaxy, originating outflows that can quench star formation in the host galaxy by heating and removing interstellar gas and dust from the ISM (i.e., AGN energy/momentum feedback). Still, these jets (that are driven by thin disk accretion) are rather ineffective in producing radio emission so that the AGN is radio quiet and does not emerge clearly from the host galaxy emission in the radio domain \citep[see][and references therein]{Mancuso2017:2017ApJ...842...95M}. This must be the case of the galaxies that perfectly overlap the galaxy main-sequence at the corresponding redshift (crf. Fig. \ref{fig_MS_evo_lines}). For a few objects we found the radio emission to be more extended than the FIR one, possibly being a signature of the forthcoming AGN feedback \citep[e.g., UDF11; see][]{Rujopakarn2016:2016ApJ...833...12R}. AGN feedback can be associate also to the broad ($\Delta v \sim500$ km/s), double-peaked CO emission line profile observed for UDF3, UDF8 and ALESS067.1, that may indicate the presence of a molecular outflow. However, we need higher resolution images to provide the right interpretation of this evidence, that is also consistent with a rotating disk of molecular gas (i.e. unresolved; Pantoni et al. in preparation). Finally, as a consequence of AGN feedback, we expect star formation to be abruptly quenched: afterwards stellar populations evolve passively and the galaxy must become a red and dead ETG.

Also the recent outcomes by \citet{Stacey2020:2020MNRAS.tmp.3229S}, studying seven gravitationally lensed quasars at $z=1.5-2.8$, support the AGN evolution as depicted by \textit{in-situ} galaxy-BH co-evolution scenario.
Interestingly, the new finding by \citet{Rizzo2020:2020Natur.584..201R} of a massive (M$_\star\simeq1.2\times10^{10}$ M$_\odot$) rotationally-supported ($v/\sigma=9.7\pm0.4$) IR-luminous (L$_{\rm IR}=(2.4\pm0.4)\times10^{12}$ L$_\odot$) galaxy at $z\sim4.2$, experiencing an intense episode of star-formation (SFR $=352\pm65$ M$_\odot$ yr$^{-1}$ and $\tau_{\rm depl}=38\pm9$ Myr), is in perfect alignment with the aforementioned scenario and with its prediction on galaxy kinematics, witnessing that the main channels of such a strong and dusty star formation activity must be \textit{in-situ}. Indeed, if the star-forming burst had been triggered by high-z wet merging events, we would have expected the galaxy to be dynamically hot, chaotic and strongly unstable: e.g., the most recent cosmological magneto-hydrodynamical simulation TNG50 gives $v/\sigma<3$ for those values of stellar mass and redshift \citep[][]{Pillepich2019:2019MNRAS.490.3196P}.
Recently, an increasing number of studies on $z\sim2$ massive star-forming galaxy kinematics have recognized the interaction/merging triggered bursts not to dominate DSFG population \citep[e.g.][]{ForsterSchreiber2006:2006ApJ...645.1062F, ForsterSchreiber2011:2011ApJ...739...45F}. A similar result is found by cosmological simulations, showing that the merger rate at the peak of Cosmic SFH (i.e., $z\sim2$, corresponding to a cosmic time of $\sim3$ Gyr) is too small to explain alone the abundant population of DSFGs observed at that epoch \citep[e.g.][]{Dekel2009:2009Natur.457..451D, Stewart2009:2009ApJ...702.1005S, Hopkins2010MNRAS.402..985H, Rodriguez-Gomez2016:2016MNRAS.458.2371R}.

Although these evidences concurrently suggest that the main mechanisms leading high-z DSFG evolution can be mostly ascribed to local, \textit{in-situ} condensation processes and our results vastly support this scenario, certainly mergings and galaxy interactions play a role in determining the evolution of galaxies, especially in galaxy (proto-)clusters and densely populated environments \citep[e.g.,][]{Tadaki2019:2019ApJ...876....1T}. 

In a forthcoming paper (Pantoni et al. in preparation) we will provide a refinement of the analysis by focusing on the ALMA view of the sample and using additional information collected from literature, such as multi-wavelength images from public catalogs. Any evidence of interaction/merging, as well as any signature of AGN feedback, will be then included in the final picture.

\section{Summary and conclusions}\label{sec_sum_conclusions}

We have presented a panchromatic study of 11 DSFGs at the peak of Cosmic SFH, that we selected in the (sub-)millimetre regime requiring the following criteria to be fulfilled for each galaxy:
3 or more detections in the optical domain ($\lambda_{\rm obs}=0.3-1$ $\mu$m);
6 or more detections in the NIR+MIR bands ($\lambda_{\rm obs}=1-25$ $\mu$m);
2 or more detections in the FIR band ($\lambda_{\rm obs}=25-400$ $\mu$m);
spectroscopically confirmed redshift in the range $1.5<z_{\rm spec}<3$;
1 or more detections and/or upper limits in the radio and X-ray regimes.

The sources are located in one of the deepest multi-band field currently available, the GOODS-S. We exploited the extensive multi-wavelength photometry, from the X-ray to the radio band, to reliably re-construct and precisely model galaxy SED, by using a physically-motivated modelling of stellar light attenuation by dust. We used CIGALE to extract the main astrophysical properties of our DSFGs (e.g., SFR, stellar mass, stellar attenuation law by dust, dust temperature, IR luminosity) from their SEDs. We exploited the Rayleigh-Jeans dust continuum to estimate galaxy dust mass (M$_{\rm dust}$) and, when CO spectroscopy was not available, their gas mass (total M$_{\rm gas,\,tot}$ and molecular M$_{\rm gas,\,mol}$). Finally, we took advantage of the X-ray and radio photometry to guess the presence of an AGN. In the following we summarize our main findings.

\begin{itemize}
    \item The 11 DSFGs of our sample are (almost) main-sequence objects, with a median M$_\star=6.5\times10^{10}$ M$_\odot$ and SFR $\sim250$ M$_\odot$ yr$^{-1}$. They are experiencing an intense and dusty (median L$_{\rm IR}\sim2\times10^{12}$ L$_\odot$) burst of star formation, with typical duration $\tau_{\star}$ ranging between 0.2 and 1 Gyr. Although their young age, the interstellar dust content is high (M$_{\rm dust}>\times10^8$ M$_\odot$), possibly due to a very rapid enrichment of the ISM (on typical timescales of $10^7-10^8$ yr). The gas mass (median M$_{\rm gas,\,tot}\sim6\times10^{10}$ M$_\odot$ and M$_{\rm gas,\,mol}\sim3\times10^{10}$ M$_\odot$), fuelling the dusty star formation, will be rapidly depleted, over a median timescale $\tau_{\rm depl}\sim200$ Myr. Nine objects out of eleven have an X-ray luminous (L$_{2-10 {\rm keV}}\gtrsim10^{42}$ erg s$^{-1}$) counterpart in the \textit{Chandra} $\simeq7$ Ms catalog and two of them are clearly dominated by the active nucleus emission (L$_{2-10 {\rm keV}}\gtrsim10^{43}-10^{44}$ erg s$^{-1}$). The radio luminosity is consistent with the emission coming from galaxy star formation, suggesting that the AGN, if present, should be radio silent or quiet.
    \item We interpret our outcomes in light of the \textit{in-situ} galaxy-BH co-evolution scenario \citep[see e.g.,][]{Mancuso2016a:2016ApJ...823..128M, Mancuso2016b:2016ApJ...833..152M, Shi2017:2017ApJ...843..105S, Mancuso2017:2017ApJ...842...95M, Lapi2018a:2018ApJ...857...22L}, that provides a possible consistent picture of high-z DSFG formation and evolution. In particular we compare our results with the predictions by the analytic model presented in \citet[]{Pantoni2019:2019ApJ...880..129P}, describing the spatially-averaged time evolution of the gas, dust, stellar and metals content in high-z star-forming counterparts of local ETGs, following the prescriptions by the aforementioned \textit{in-situ} scenario. We find our outcomes to match the model predictions within their $2\sigma$ scatter, suggesting that the main drivers of the evolution of our 11 DSFGs can be traced back mostly to local condensation processes.
    \item We complemented these results by exploiting multi-wavelength images from public catalogs, that allowed us to include in our final interpretation every signature of galaxy merging/interactions and feedback. The compact FIR and radio sizes ($\lesssim$ a few kpc) of our DSFGs, together with their optical radii ($\sim2-6$ kpc), suggest that the bulk of their star formation can be traced back to in-situ condensation processes. Most of our sources shows an optical isolated morphology, while four galaxies (UDF11, ALESS067.1, AzTEC.GS21, AzTEC.GS22) have more complex (i.e., clumpy) morphologies , possibly indicating the presence of minor companions - that can prolong the star formation in the dominant galaxy by refuelling it with gas - or just being a signature of the ongoing dusty star formation. Higher resolution imaging are needed to definitely clarify the picture. However, we do not expect these interactions to have an important impact on the subsequent evolution of the dominant galaxy.
    \item Following the predictions by the \textit{in-situ} galaxy-BH co-evolution \citep[see][]{Lapi2018a:2018ApJ...857...22L}, we can state that the majority of the galaxies in our sample is caught in the compaction phase and we expect them to be quenched by the AGN feedback in $\lesssim10^8$ yr. For four objects we found some signatures of AGN feedback either in the radio band (UDF11), that appears more extended than the FIR one, or by the detection of possible AGN-driven molecular outflows (UDF3, UDF8, ALESS067.1). We expect their subsequent evolution to be passive, mainly driven by their stellar populations aging and mass additions by dry merger events, and ultimately to become compact quiescent galaxies or massive ETGs. In a forthcoming paper (focusing on the ALMA view of the sample; Pantoni et al. in preparation) we will gathered together all these evidences in order to provide a novel approach in characterizing the individual DSFG and predicting its subsequent evolution.
    \item We have compared the results obtained for our sample of 11 spectroscopically confirmed $z\sim2$ DSFGs with other recent good studies on high-z DSFGs exploiting SED fitting, and we find a great agreement between the median values of the main physical quantities estimated for these galaxies, such as stellar mass, gas mass and dust mass. Thus, we can conclude that our approach and selection criteria do not introduce any substantial bias and may be applied to statistical samples of spectroscopically-confirmed DSFGs, as soon as they will be available (see e.g., the ongoing z-GAL NOEMA Large Program; PIs: P. Cox; T. Bakx; H. Dannerbauer).
    \item Finally, we would highlight the importance of combining multi-band photometry, gas spectroscopy and high-resolution imaging with a physically motivated model in order to characterize the role of high-z DSFGs in the context of galaxy formation and evolution and the impact of galaxy interactions and AGN feedback in determining their evolution. 
\end{itemize}

\section*{Acknowledgements}
The authors thank the anonymous referee for stimulating and constructive comments that helped to improve this study. LP gratefully thanks S. Campitiello for the helpful discussions.

This paper makes use of the following ALMA data: ADS/JAO.ALMA\#2011.0.00294.S (PI: Smail); \#2012.1.00983.S (PI: Leiton); \#2012.1.00173.S (PI: Dunlop); \#2015.1.00098.S (PI: Kohno); \#2015.1.00543.S (PI: Elbaz); \#2015.1.00948.S (PI: da Cunha); \#2015.1.01074.S (PI: Inami); \#2015.1.00242.S and \#2016.1.01079.S (PI: Bauer); \#2016.1.00564.S (PI: Weiss); \#2017.1.01347.S (PI: Pope). ALMA is a partnership of ESO (representing its member states), NSF (USA), and NINS (Japan), together with NRC (Canada), NSC and ASIAA (Taiwan), and KASI (Republic of Korea), in cooperation with the Republic of Chile. The Joint ALMA Observatory is operated by ESO, AUI/NRAO, and NAOJ. We acknowledge financial support from the grants: PRIN MIUR 2017 prot. 20173ML3WW 001 and PRIN MIUR 2017 prot. 20173ML3WW 002 (‘Opening the ALMA window on the cosmic evolution of gas, stars and massive black holes’).
A. Lapi is supported by the EU H2020-MSCAITN-2019 Project 860744 ‘BiD4BEST: Big Data applications for Black hole Evolution STudies’. 

\section*{Data Availability}
 
This article uses public data products from ALMA Archive (repository available at the following link: https://almascience.nrao.edu/asax/; note that project codes of interest are listed in the Acknowledgements).

Photometric optical, infrared, radio and X-ray data come from (in the order): 
\begin{itemize}
    \item GOODS-MUSIC sample: a multi-colour catalog of near-IR selected galaxies in the GOODS-South field \citep[][available at the link: https://cdsarc.unistra.fr/viz-bin/cat/J/A+A/449/951; VizieR DOI: 10.26093/cds/vizier.34490951]{Grazian06:2006yCat..34490951G};
    \item combined PEP/GOODS-Herschel data of the GOODS fields by \citet{Magnelli2013:2013A&A...553A.132M} at the link https://www.mpe.mpg.de/ir/Research/PEP/DR1 and publicly available at\\
http://www.mpe.mpg.de/ir/Research/PEP/public\_data\_releases.php \citep[see also][available at the link: https://cdsarc.unistra.fr/viz-bin/cat/J/A+A/528/A35; VizieR DOI: 10.26093/cds/vizier.35280035]{Magnelli11:2011yCat..35280035M};
\item Herschel Multi-tiered Extragalactic Survey: HerMES \citep[][publicly available through the Herschel Database in Marseille, HeDaM, at http://hedam.oamp.fr/HerMES and at the VizieR links https://cdsarc.unistra.fr/viz-bin/cat/VIII/95 and https://cdsarc.unistra.fr/viz-bin/cat/VIII/103 ]{Oliver14:2014yCat.8095....0O, Oliver17:2017yCat.8103....0H};
\item Very Large Array 1.4 GHz survey of the Extended Chandra Deep Field South: second data release \citep[][available at the link: https://cdsarc.unistra.fr/viz-bin/cat/J/ApJS/205/13; VizieR DOI : 10.26093/cds/vizier.22050013]{Miller13:2013yCat..22050013M};
\item Very Large Array 6 GHz imaging by \citet{Rujopakarn2016:2016ApJ...833...12R} \citep[follow-up of the B6 ALMA sources by][in the HUDF-S; project ID ADS/JAO.ALMA\#2012.1.00173.S]{Dunlop2017:2017MNRAS.466..861D}: project ID VLA/14A-360.
\item Chandra Deep Field-South survey: 7 Ms source catalogs \citep[][available at the VizieR link: https://cdsarc.unistra.fr/viz-bin/cat/J/ApJS/228/2]{Luo17:2017yCat..22280002L}.
\end{itemize}



\bibliographystyle{mnras}
\bibliography{main} 


\appendix

\section{Derivation of individual galaxy attenuation law}\label{appendix_attlaw}

We derive the individual galaxy attenuation law (shown in Fig. \ref{panel}, right column), starting from stellar luminosities calculated by CIGALE. In particular, we describe how the two dust components (populating galaxy ISM and BCs) concurrently draw galaxy total attenuation law.

By definition, the attenuation law is a function of wavelength normalized to attenuation in the V (photometric) band, i.e. $A_\lambda/A_V$.
Namely, a non-transparent medium along the line-of-sight (los) modify the source intrinsic monochromatic luminosity $L_\lambda^0$ according to the law:
\begin{equation}\label{L}
 \frac{L_\lambda}{L_\lambda^0}=e^{-\tau_\lambda}
\end{equation}
where $\tau_\lambda$ is the medium optical depth and $L_\lambda$ is the attenuated luminosity at a given $\lambda$.
The attenuation at a given wavelength $\lambda$ is then defined as:
\begin{equation}\label{A}
 A_\lambda = -2.5 \log_{\rm 10} \frac{L_\lambda}{L_\lambda^0}.
\end{equation}
Comparing equations (\ref{L}) and (\ref{A}), it follows that $A_\lambda = 2.5\log_{\rm 10}(e) \,\tau_\lambda = 1.086 \,\tau_\lambda \simeq \tau_\lambda$. 

We note that $A_\lambda$ is actually a measure of the stellar luminosity which has been absorbed by dust, i.e $\Delta L_\lambda= L_\lambda^0-L_\lambda$. In our case, $\Delta L_\lambda^{\rm BC}$,  $\Delta L_\lambda^{\rm ISM}$ and  $\Delta L_\lambda^{\rm tot}$ are provided by CIGALE for every galaxy in our sample, as outputs of SED fitting, together with stellar intrinsic luminosities, thus: 
\begin{equation}\label{A_L}
 A_\lambda^{\rm i} = -2.5 \log_{\rm 10} \frac{L_\lambda^{0,\,i}-\Delta L_\lambda^{\rm i}}{L_\lambda^{0,\,i}}.
\end{equation}
where $i=$ BC, ISM, tot.

Attenuations in V-band ($A_V^i$) are obtained from Eq.~\eqref{A_L}, taking the values corresponding to $\lambda = 5500$ \AA{}.
Then, deriving attenuation laws $A_\lambda^i/A_V^i$ is straightforward.

\section{Derivation of galaxy intrinsic 2-10 keV luminosity}\label{appendix_luo}

In the following we convert the $0.5-7.0$ keV luminosity by \citet{Luo2017:2017ApJS..228....2L} to the corresponding $2-10$ keV luminosity, given that the latter is the most widely used in literature to investigate galaxy evolution. 
We exploited the known intrinsic spectral index \citep[$\equiv1.8$ or greater;][]{Luo2017:2017ApJS..228....2L} to estimate the conversion factor by using the X-ray simulator WebPIMMS\footnote{https://heasarc.gsfc.nasa.gov/cgi-bin/Tools/w3pimms/w3pimms.pl}, the same tool used by \citet{Luo2017:2017ApJS..228....2L} to obtain N$_{\rm H,int}$. Intrinsic and effective\footnote{Note that the effective spectral index is actually almost coincident with the observed spectral index.} spectral indices ($\Gamma_{\rm int}$ and $\Gamma_{\rm eff}$), together with $L_{0.5-7\,\rm keV} - L_{2-10\,\rm keV}$ conversion factors and the derived $2-10$ keV intrinsic luminosities ($L_{2-10\,\rm keV}$) are listed in Tab. \ref{tab_luo}.

Then, we convert the $2-10$ keV luminosities to be consistent with the redshift adopted in this work (crf. Tab. \ref{tab_redshifts_coords}), not always exactly coincident with \citet{Luo2017:2017ApJS..228....2L} ones. To this aim, we proceed as follows. If both the intrinsic X-ray luminosity $L_{\nu/(1+z)}$ of a source at a given redshift $z$ and the power-law describing its X-ray 
emission are known, then it is possible to derive the corresponding observed flux $S_{\nu/(1+z)}$ applying the so-called {\itshape k-correction}.
This correction is defined by the following equation:
\begin{equation}\label{eq_k_corr}
 S_{\nu/(1+z)}=(1+z)\,\frac{L_\nu}{L_{\nu/(1+z)}}\,\frac{L_{\nu/(1+z)}}{4\pi\,{d_{\rm L}(z)}^2}
\end{equation}
Since $L_\nu\propto\nu^{-\alpha}$, with $\alpha>0$ and $\Gamma=\alpha+1>0$ (e.g., Ishibashi \& Courvoisier 2010) where $\Gamma$ is the energy spectral index,
one can write:
\[
 \frac{L_\nu}{L_{\nu/(1+z)}}=\Biggl(\frac{\nu}{\nu/(1+z)}\Biggr)^{-\alpha}=(1+z)^{-\alpha}=(1+z)^{1-\Gamma}
\]
this result, together with equation (\ref{eq_k_corr}), brings to:
\begin{equation}\label{k-correction_X}
 S_{\nu/(1+z)}=(1+z)^{2-\Gamma_{\rm int}}\,\,\frac{L_{\nu/(1+z)}}{4\pi\,{d_{\rm L}(z)}^2}
\end{equation}
where $\Gamma_{int}$ is the intrinsic energy spectral index, i.e. the one derived after the correction for an eventual intrinsic absorption.
Specializing Equation (\ref{k-correction_X}) to our case, we have:
\begin{equation}\label{k-correction_luo}
 S_{2-10\,\rm keV}=(1+z_{\rm L})^{2-\Gamma_{\rm int}}\,\,\frac{L_{2-10\,\rm keV}(z_{\rm L})}{4\pi\,{d_{\rm L}(z_{\rm L})}^2}
\end{equation}
where $z_{\rm L}$ is the source redshift in \citet{Luo2017:2017ApJS..228....2L}. Then, exploiting equation (\ref{k-correction_luo}) it is possible to derive the observed flux $S_{2-10\,\rm keV}$ corresponding to the intrinsic luminosity $L_{2-10\,\rm keV}(z_{\rm L})$ provided by \citet{Luo2017:2017ApJS..228....2L}. Now, it is sufficient to invert equation (\ref{k-correction_X}) with $S_{\nu/(1+z)}\equiv S_{2-10\,\rm keV}$ and from equation (\ref{k-correction_luo}) it follows:
\begin{equation}
 L_{X,\,\rm int}=(1+z_{\rm L})^{2-\Gamma_{\rm int}}\,\,\frac{L_{2-10\,\rm keV}(z_{\rm L})}{4\pi\,{d_{\rm L}(z_{\rm L})}^2}\,\frac{4\pi\,{d_{\rm L}(z)}^2}{(1+z)^{2-\Gamma_{\rm int}}}\,.
\end{equation}
Intrinsic $2-10$ keV luminosities ($L_{X,\,int}$) are listed in Tab. \ref{tab_luo}. The other quantities exploited to derive $L_{X,\,int}$ are listed in Tab. \ref{tab_luo_appendix}.

\begin{table*}
\centering
 \caption{In this Tab. we list: IDs of the source associations (ID: this work; ID$_{\rm X}$: Luo et al. 2017) and their angular separation (d) in arcsec; redshifts adopted in this work (z) and the ones associated to X-ray sources in \citep[][, $z_{\rm L}$]{Luo2017:2017ApJS..228....2L}; intrinsic $0.5-7$ keV luminosities from Luo et al. (2017, $L_{0.5-7 \rm keV}$); conversion factors (f$_{\rm conv}$) from $L_{0.5-7\,\rm keV}$ to $L_{2-10\,\rm keV}$; $2-10$ keV luminosities at redshift $z_{\rm L}$ ($L_{2-10\,\rm keV}$); $2-10$ keV intrinsic luminosities at redshift $z$ ($L_{\rm X}$); effective spectral indices ($\Gamma_{\rm eff}$) and the intrinsic ones ($\Gamma_{\rm int}$). In the last two columns we show (in the order): the class (AGN or galaxy) associated to each source by \citet{Luo2017:2017ApJS..228....2L} and the X-ray dominant component (active nucleus or host galaxy) found by our analysis.}\label{tab_luo_appendix}
  \begin{tabular}{lcccccccccc}
 \hline
 \bfseries {ID} & $\mathbf{ID_X}$ & \bfseries {d} &$\mathbf{z}$ & $\mathbf{z_{L}}$ & $\mathbf{L_{0.5-7\,keV}}$  & $\mathbf{f_{conv}}$ & $\mathbf{L_{2-10\,keV}}$ & $\mathbf{\Gamma_{eff}}$ & $\mathbf{\Gamma_{int}}$ \\
  & & [arcsec] & & & [$10^{42}$ erg s$^{-1}$]  &  & [$10^{42}$ erg s$^{-1}$]  &  & \\
 \hline
 UDF1  & 805 & 0.69 & 2.688 & 2.69 & 64.0 & 0.63 & 40.3 &1.96 &1.96 \\
  UDF3    & 718 & 0.54 & 2.544 & 2.547& 4.6 & 0.40 & 1.8 &2.44 &2.44  \\
  UDF8    & 748 & 0.07& 1.549 & 1.552 &50.7 &0.72  & 36.5 &1.32&1.8 \\
  UDF10 & 756 & 0.31 & 2.086 & 2.096  &2.7  & 0.22 &  0.6& 3.0&3.0\\
  UDF11  & 751 & 0.29 & 1.996 & 1.998 & 2.4 & 0.72 & 1.7& 1.74&1.8\\
  UDF13  & 655 & 0.26 & 2.497 & 2.07 & 2.3 & 0.57 &  1.3& 2.07& 2.07\\
  ALESS067.1  & 794 & 0.40 & 2.1212 & 2.122 & 8.5 & 0.45 &  3.8 &2.33&2.33\\
  AzTEC.GS25  & 844 & 0.71 & 2.292 & 2.292 & 8.5 &0.72  & 6.1 & 1.2&1.8\\
  AzTEC.GS21 & 852 & 0.36 & 1.91 & 1.91 & 2.3 &0.72 & 1.7 & 1.4& 1.8\\
  \hline
  \end{tabular}
\end{table*}


\bsp	
\label{lastpage}
\end{document}